\documentclass[showkeys, pra,aps,twocolumn,twoside,showpacs]{revtex4-2}
\usepackage[T1]{fontenc}
\usepackage{amsfonts}
\usepackage{amssymb}
\usepackage{mathrsfs}
\usepackage[intlimits]{amsmath}
\usepackage{bm, bbm}
\usepackage[dvips]{graphicx}

\usepackage[T1]{fontenc}
\usepackage[latin2]{inputenc}
\usepackage {hyperref}
\usepackage{amsmath}
\usepackage{latexsym}
\usepackage{soul,xcolor}
\usepackage{color}
\usepackage{graphicx}
\usepackage[export]{adjustbox}
\usepackage{subcaption}
\usepackage{float}

\def\be{\begin{equation}}
\def\ee{\end{equation}}
\def\ben{\begin{eqnarray}}
\def\een{\end{eqnarray}}
\newcommand{\ket}[1]{| #1 \rangle}
\newcommand{\bra}[1]{\langle #1 |}

\DeclareMathOperator{\sinc}{sinc}

\begin{document}
\setstcolor{red}
\title{Encoding position by spins: Objectivity in the boson-spin model}

\author{Tae-Hun Lee}\email{taehunee@cft.edu.pl }
\affiliation{Center for Theoretical Physics, Polish Academy of Sciences,\\ Aleja Lotnik\'ow 32/46, 02-668 Warsaw, Poland}
\author{Jaros\l{}aw K. Korbicz}\email{jkorbicz@cft.edu.pl}
\affiliation{Center for Theoretical Physics, Polish Academy of Sciences,\\ Aleja Lotnik\'ow 32/46, 02-668 Warsaw, Poland}
\date{\today}

\begin{abstract}
We investigate quantum objectivity in the boson-spin model, where a central harmonic oscillator interacts with a thermal bath of spin-$\frac{1}{2}$ systems. We analyze how information about a continuous position variable can be encoded into discrete, finite-dimensional environments. More precisely, we study conditions under which the so-called spectrum broadcast structures (SBS) can be formed in the model. These are multipartite quantum state structures, representing a mode-refined form of decoherence. Working in the recoil-less limit, we use the Floquet theory to show that despite its apparent simplicity, the model has  a rich structure with different regimes, depending on the motion of the central system. In one of them, the faithful encoding of the position and hence objectivity are impossible irrespectively of the resources used. In another, large enough collections of spins will faithfully encode the position information. We derive the characteristic length scales, corresponding to decoherence and precision of the encoding.
\end{abstract}


\maketitle
\section{Introduction}
 Although quantum mechanics is believed to be the most fundamental theory of Nature, it is mysterious and puzzling that we still have not fully succeeded in explaining our daily observed world by quantum mechanics. How is it possible for all counter-intuitive quantum natures like superposition, interference, disturbance, non-locality, etc., clearly disappear in our macroscopic world? Although many alternative approaches compete for explaining the quantum-to-classical transition, none of them has been agreed on so far. In this situation it is important to restrict ourselves to the question how far the classical-quantum discrepancy can be explained within the current-state quantum mechanics. One of the aspects of the problem is the objective character of the macroscopic world as first noted by Zurek \cite{ZurekNature2009, ZurekPhysToday2014}. Objectivity may be viewed as an observer-independence, bears some resemblance to the relativity theory. But due  to the inevitable disturbances introduced by observations in quantum mechanics, it is not \textit{a priori} clear how to achieve the observer independence, at least at the basic level of measurement results.  

In the history of physics, there has been an orthodox view that the world existing outside of the system of interest plays  a role of a source of noise which can be, at least in principle, perturbatively controlled, so when it is minimized, the ``true nature'' will get more and more approachable. But quantum mechanics changed that view: our macroscopic reality can be considered the consequence of interaction between a system and the rest of the world, as emphasized by the decoherence theory \cite{JoosBook2002, Schlosshauer2007}. Within this view there has been an idea developed, called quantum Darwinism, aiming at explaining the apparent observer-independence in the macroscopic world \cite{Zurek:2003zz, ZurekNature2009, ZurekPhysToday2014, Korbicz:2020tkf}. It postulates that interactions between a system and an environment redundantly transfer the information of a system to the environment during decoherence. The idea has opened a new field of objectivity studies (see, e.g., \cite{LePRA2020, CakmakEntropy2021, TouilPRL2022, BaldijaoPRXQ2022, ChisholmQuantum2023} for some of the most recent developments). Although quantum Darwinism does not completely explain the non-unitary collapse process, the famous measurement problem, it is still remarkable that within quantum mechanics some form of objective classicality can be derived.

A further development of the quantum Darwinism idea is represented by spectrum broadcast structures (SBS) \cite{Korbicz:2014, Horodecki:2015, Korbicz:2020tkf}, which are specific quantum-state structures, encoding an operational form of objectivity.  SBS are a stronger form of quantum Darwinism in a sense that SBS formation implies the original quantum Darwinism conditions but not vice versa \cite{LePRA2020}. Under appropriate conditions, SBS have been shown to form in almost all the canonical decoherence models \cite{Schlosshauer2007}, i.e., a collisional decoherence \cite{Korbicz:2014}, quantum Brownian motion (QBM) in the recoil-less limit \cite{Tuziemski:2015,Lee:2023qrm}, a spin-spin model \cite{Mironowicz:2017, Mironowicz:2018}, and a spin-boson model \cite{Lampo:2017,Tuziemski:2015}. The only one left is a boson-spin model, which we analyze in this work. The central system is a massive oscillator interacting with a thermal bath of spin-$\frac{1}{2}$ systems. We use here a recoil-less approximation, similarly as in the QBM studies \cite{Tuziemski:2015,Lee:2023qrm}, where a harmonic oscillator influences the spin environment, but the recoil is suppressed. This approximation leads to the same form of Hamiltonian as for, e.g., a two-level atom interacting with linearly polarized light \cite{Shirley:1965}. As the effective Hamiltonian for the spin environment is time periodic, this allows us to use the Floquet theory and the high-frequency expansion. The most interesting question arising in this model, and absent in previous studies, is how finite-dimensional environments encode a continuous variable (the central oscillator's amplitude). We show that depending on the state of motion of a central oscillator there can be either only a momentary formation of SBS states or a permanent or asymptotic one. 
Interestingly, this  behavior is opposite to the QBM model, showing once again the stark difference between spin and oscillator environments. We derive the length scales corresponding to decoherence and faithful information encoding in the environment, which scale $1/\sqrt{N}$, where $N$ is the environment fractions size that is used to store the information or decohere the system.

We will be interested in a so-called partially reduced state $\rho_{S:oE}$, where a fraction of the environment, assumed unobserved and denoted by $uE$, was traced out as an unavoidable loss, but the remaining fraction, denoted by $oE$, is kept for observation.  The spectrum broadcast structures (SBS) or an objective quantum state, is then defined as follows \cite{Korbicz:2014, Korbicz:2020tkf}:
\begin{align}
\rho_{S:fE}=\sum_ip_i|i\rangle_S\langle i|\otimes\rho^{E_1}_i\otimes\cdots\otimes\rho^{E_{fN}}_i,\label{SBS}
\end{align}
where 
\be
\rho^{E_k}_i\rho^{E_k}_j=0 \label{ort}
\ee
for $i\neq j$, which is equivalent to the states $\rho^{E_k}_i$ having orthogonal supports and thus being perfectly distinguishable.
After unobserved degrees of freedom traced out, the SBS of a total density matrix are in an orthogonal convex combination form. The approach to the SBS is marked by vanishing quantum coherence (off-diagonal elements) and a perfect distinguishability (diagonal elements), corresponding to a vanishing decoherence factor and a vanishing generalized overlap (state fidelity), respectively \cite{Mironowicz:2017}. These are the objectivity markers that we will analyze in various regimes.

 \section{Dynamics of System}\label{Dynamics}
 The total Hamiltonian $H$ for a simple harmonic oscillator, bilinearly interacting with spin-$\frac{1}{2}$ environment \cite{Schlosshauer2007} is given by
 \begin{align}
H=H_{S}+\sum_iH^{(i)}_{E}+\sum_iH^{(i)}_{\text{int}},\label{H}
 \end{align}
 where
 \begin{align}\nonumber
    H_{S}&=\frac{\hat{P}^2}{2M}+\frac{1}{2}M\Omega^2 \hat{X}^2,\\\nonumber
    H^{(i)}_E&=-\frac{\Delta_i}{2}\sigma^{(i)}_x,\\
    H^{(i)}_{\text{int}}&= g_i\hat{X}\otimes \sigma^{(i)}_z,\label{H_S,H_E,H_int}
 \end{align}
 where $M$ and $\Omega$ are a mass and an angular frequency of an oscillator, respectively, and $g_i$ and $\Delta_i$ are a spin-environmental coupling constant and a self-energy (also called the tunneling matrix element) for the $i$th spin system, respectively. Here only bipartite interactions $H^{(i)}_{\text{int}}$ between the $i$th spin and the harmonic oscillator are considered, without mutual interactions among the spins. Despite its simple form, the total Hamiltonian (\ref{H}) is difficult to solve directly. For the purpose of our analysis it is, however, enough to use the so-called recoil-less limit, at least as a first approximation. In this limit, the central oscillator is assumed to be massive enough not to feel the recoil of the environment, while each of the environmental spins is affected by the motion of the central oscillator, which acts as a classical force. This is an opposite limit to the much more popular Born-Markov limit, where it is an environment that is assumed not to be affected by the system. The justification for such a choice comes from the fact that we are primarily interested in an information leakage from a system to an environment as cutting the influence of the system on the environment would also cut the information leakage. Hence, it leads us to a study of the opposite, recoil-less limit. It can be viewed as a version of the Born-Oppenheimer approximation \cite{Born:1927} and it was already used in the objectivity studies in \cite{qbm_Zurek,Augusto,Tuziemski:2015}. In the recoil-less limit, the system $S$ evolves unperturbed, according to its own dynamics $H_S$. It influences the environment via the interaction Hamiltonian where the system's position operator $\hat{X}$ can be approximated by the classical trajectory $X(t;X_0)$, starting at some initial position $X(0)=X_0$.  The resulting approximate solution is given by the following ansatz \cite{Tuziemski:2015}:  
\begin{equation}\label{final}
\ket{\Psi_{S:E}}=\int dX_0 \phi_0(X_0) e^{-i\hat H_St}|X_0\rangle \hat U_{\text{eff}}(X(t;X_0))\ket{\psi_0},
\end{equation} 
Here $\hbar=1$ convention has been used and will be applied to the entire presentation.
$\hat U_{\text{eff}}(X(t;X_0))$ is the evolution generated by 
\begin{align}
     H_{\text{eff}}&=\sum_i\left(-\frac{\Delta_i}{2}\sigma^{(i)}_x+g_iX(t;X_0)\sigma^{(i)}_z\right),\label{Heff}
 \end{align}
and $\ket{\phi_0}$ and $\ket{\psi_0}$
are initial states of $S$ and $E$ respectively. Formally, (\ref{final}) is generated by a controlled-unitary evolution:
\begin{equation}\label{USE}
\hat U_{S:E}(t)=\int dX_0 e^{-i\hat H_St}\ket{X_0}\bra{X_0}\otimes U_{\text{eff}}(X(t;X_0)),
\end{equation}
acting on the initial state $\ket{\phi_0}\ket{\psi_0}$. 
For simplicity,  we will limit ourselves to trajectories obtained when the system is initially in the squeezed coherent state in position 
$\ket{\phi_0}=\hat{D}(\alpha)\hat S(r)\ket{0}$, where $\hat{D}(\alpha)=e^{\alpha \hat{a}^\dagger-\alpha^*\hat{a}}$ and $\hat S(r)\equiv e^{r(\hat a ^2-\hat a^{\dagger 2})/2}$.
First, we may assume that initial position is the maximum $X_0$ with the squeezed vacuum wave function in position, so that:
\be\label{traj}
X(t;X_0)=X_0\cos(\Omega t). 
\ee
For the analysis of a zero initial position, we may assume $X(t=0)=0$ and take $X(t;X_0)=X_0\sin(\Omega t)$.

Assuming a completely separable initial state:
\be\label{prod} 
\rho_{S:E}(0)=\rho_{S}(0)\otimes \bigotimes_i \rho_{E}^{(i)}(0),
\ee
which is motivated here by the fact that we wish to study buildup of the system-environment correlations, the full solution is easily obtained from (\ref{USE}): 
\begin{align}\label{rho_SE}
   \rho_{S:E}(t)&= \int dX_0dX'_0 \rho(X_0, X'_0) e^{-iH_St}|X_0\rangle\langle X_0'| e^{iH_S t}\nonumber\\
& \otimes\bigotimes_{i=1}^N U_i(X_0,t)\rho^{(i)}_E(0) U^\dagger_i(X'_0,t),
\end{align}
where
\be
\rho(X_0, X'_0)\equiv\langle X_0|\rho_S(0)|X'_0\rangle
\ee
are the initial coherences and the conditional evolutions of the $i$th spin, $U_i(X_0,t)$, are generated by:
\be\label{Hi}
H_i=-\frac{\Delta_i}{2}\sigma^{(i)}_x+g_iX_0\cos(\Omega t)\sigma^{(i)}_z.
\ee
This allows us to find the effective evolution of spin states. The above Hamiltonian is well known and describes, 
e.g., an interaction of a linearly  polarized light with a two-level atom \cite{Shirley:1965}.  The periodicity in time allows us to use the standard methods of the Floquet theory (see  \cite{Floquet:1883} for the historical work and, e.g., \cite{Rahav:2003, Goldman:2014} for modern expositions) to find approximate solutions. 

The Floquet theorem states that a unitary evolution for a periodic Hamiltonian can be written as a product of a unitary evolution driven by a periodic time-dependent Hamiltonian $K(t)$ with the same period of the Hamiltonian and a unitary evolution by a time-independent Hamiltonian $H_{F}$:
\begin{align}
    U(t,t_0)&=e^{-iK(t)}e^{-i(t-t_0)H_{F}}e^{iK(t_0)}.\label{Floquet_U}
\end{align}
$H_{F}$ is responsible for a slow dynamics forming an overall profile while $K(t)$ for a fast dynamics forming internal profile with the same oscillator periodicity $K(t)=K(t+T)$ as the given periodic Hamiltonian (\ref{Hi}). The operators $K(t)$ and $H_F$ can be perturbatively identified by using the high-frequency expansion \cite{Rahav:2003, Goldman:2014}. One Taylor expands $H_F$ in $1/\Omega\ll1$, while $K(t)$ is the remaining part after  $H_{F}$ has been taken out.
In general, convergence of the expansion is not always guaranteed \cite{Rahav:2003,Goldman:2014}. By imposing conditions on the rest of parameters with $\Omega$, the convergence of the series can be controlled, as we remark below. 
Fourier expanding the Hamiltonian (\ref{Hi}),
\begin{align}
    H_i=H_0+\sum^\infty_{j=1}(V^{(j)}e^{ij\Omega t}+V^{(-j)}e^{-ij\Omega t}),\label{Hi}
\end{align}
where $H_0=-(\Delta/2)\sigma_x$ and $V^{(1)}=V^{(-1)}=(gX_0/2)\sigma_z$, the rest being zero, one finds the high-frequency expansion of $H_F$ and $K(t)$ (we omit the environment index $i$ for simplicity):
\begin{align}\nonumber
    H_{F}&=H_0+\frac{1}{\Omega}\sum^\infty_{j=1}\frac{1}{j}[V^{(j)},V^{(-j)}]\\
    &+\frac{1}{2\Omega^2}\sum^\infty_{j=1}\frac{1}{j^2}([[V^{(j)},H_0],V^{(-j)}]+\text{H.c.})+\cdots,\label{def:He}
\end{align}
and
\begin{align}\nonumber
    K(t)&=\frac{1}{i\Omega}\sum^\infty_j\frac{1}{j}(V^{(j)}e^{ij\Omega t}-V^{(-j)}e^{-ij\Omega t})\\
    &+\frac{1}{i\Omega^2}\sum^\infty_j\frac{1}{j^2}([V^{(j)},H_0]e^{ij\Omega t}-\text{H.c.})+\cdots.\label{def:K}
\end{align}
For the purpose of this work, we take the lowest non-vanishing correction terms only and $t_0=0$. This gives:
\begin{align}
    tH_{F}
      &=-\tilde{\Delta}(1-\xi^2)\tau\sigma_x,\label{He}\\
    K(t)&=\xi\sigma_z\sin\tau\label{K}
\end{align}
where we introduced dimensionless position $\xi$, tunneling energy $\tilde{\Delta}$, and time $\tau$:
\begin{align}\label{parameters}
\xi\equiv\frac{gX_0}{\Omega}, \ \tilde{\Delta}\equiv\frac{\Delta}{2\Omega},\ \tau\equiv \Omega t.
\end{align}
Picking the initial time $t_0=0$, there is no initial kick $K(0)=0$. The convergence of the expansion is guaranteed for $\tilde{\Delta}, \xi\ll1$.
Using \eqref{He} and \eqref{K}, the unitary evolutions defined in (\ref{Floquet_U}) are easily found:
\begin{align}\nonumber
 &e^{-i(t-t_0)H_{F}}=\cos[\tilde{\Delta}(1-\xi^2)\tau]+i\sigma_x\sin[\tilde{\Delta}(1-\xi^2)\tau],\\
&e^{-iK(t)}=\cos[\xi\sin\tau]-i\sigma_z\sin[\xi\sin\tau].\label{U_High_Frequency}
\end{align}
The first is the slow motion part of the dynamics, while the second is the fast motion part (the micromotion),
with the time-periodic frequency proportional to $\sin\tau$.  
They lead to the following effective evolution, modulo $O(\Omega^{-2})$ terms:
\begin{align}\label{Uk}
U_k(X_0,t)&=\left[\cos(\xi_k\sin\tau)-i\sigma^{(k)}_z\sin(\xi_k\sin\tau)\right]\\\nonumber
&\times \left[\cos[\tilde{\Delta}_k(1-\xi_k^2)\tau]+i\sigma^{(k)}_x\sin[\tilde{\Delta}_k(1-\xi_k^2)\tau]\right]. 
\end{align}

\section{Objective Quantum States}
We now investigate possibilities of a formation of the SBS-like state (\ref{SBS}). The form of the evolution (\ref{USE}) dictates that the corresponding pointer state eigenvalues are the initial oscillator position $X_0$, equivalently its amplitude, that controls the evolution of the environment and hence leaks into it. 
Following the general quantum Darwinism setup, we assume that part of the environment, called $oE$, is under observation while the rest, called $uE$, is unobserved. 
We are thus interested in the partial trace of (\ref{rho_SE}) over $uE$: 
\begin{align}\label{rho_SoE}
 & \rho_{S:oE}(t)=\text{Tr}_{uE}\rho_{S:E}(t)\\\nonumber
&=\int dX_0dX'_0 \rho(X_0, X'_0) \Gamma_{X_0,X'_0} e^{-iH_St}|X_0\rangle\langle X_0'| e^{iH_S t} \\\nonumber
&\otimes\bigotimes_{i \in oE} U_i(X_0,t)\rho^{(i)}_E(0) U^\dagger_i(X'_0,t),
\end{align}
where
 \begin{align}\label{Gamma}
    \Gamma_{X_0,X'_0}&\equiv\prod_{k\in uE}\text{Tr}[U_k(X_0,t)\rho^{(k)}_E(0)U^\dagger_k(X'_0,t)]  \\
    &= \prod_k\Gamma^{(k)}_{X_0,X'_0} \nonumber
\end{align}
is the decoherence factor, associated with the unobserved part of the environment $uE$. 
We note that since the decoherence factor is a function, which magnitude is always less than one, it can never be $\sim\delta(X_0-X'_0)$, and hence a full decoherence and a strict $S:E$ disentanglement cannot happen, though they take place for discrete variables, but they happen rather in an existence of some decoherence length, below which coherences are preserved \cite{Kittelbook, Lee:2023qrm}. 

To identify under such conditions the candidates for the information-encoding states $\rho^{E_k}_i$ from (\ref{SBS}), we recall that in the Darwinism setup, the observers  monitor the system only indirectly, via portions of the environment. Since in realistic conditions, a single environment will in general carry a vanishingly small information about the system, we assume that each observer has an access to a collection of environments, called macrofraction \cite{Korbicz:2014}, scaling with the total number of environments $N$. The state of a macrofraction is obtained from (\ref{rho_SoE}) by tracing out everything except for the given macrofraction:
\label{rho_mac}\begin{align}
 & \rho_{\text{mac}}(t)=\text{Tr}_{S oE \setminus \text{mac}}\rho_{S:oE}(t)\\\nonumber
&=\int dX_0 p(X_0) \rho_{\text{mac}}(X_0),
\end{align}
where $p(X_0)\equiv \langle X_0|\rho_S(0)|X_0\rangle$ is the probability distribution of the initial position and:
\be
 \rho_{\text{mac}}(X_0)\equiv \bigotimes_{k \in \text{mac}} U_k(X_0,t)\rho^{(k)}_E(0) U^\dagger_k(X_0,t)\label{mac}
\ee
is the conditional state corresponding to $X_0$. Thus, to know $X_0$, each observer must be able to distinguish the states (\ref{mac}) for different $X_0$. There are different scenarios to study state distinguishability \cite{Fuchs:1997ss}, e.g., quantum Chernoff bound used already in some studies of quantum Darwinism \cite{Chernoff:1952,Zurek:2016} or, more appropriate here due to the continuous parameter, quantum metrology \cite{Braunstein:1994zz}. Here, like in the the previous SBS studies, we will follow a simple form of the  latter and will study the generalized overlap (state fidelity), which is the integral of the Quantum Fisher Information \cite{Helstrom:1969fri}:
\begin{align}
B(\rho,\rho')\equiv\left[\text{Tr}\sqrt{\sqrt{\rho}\rho'\sqrt{\rho}}\right]^2.\label{B}
\end{align}
Note $B(\rho,\rho')$ here is defined by squaring, which is different from the usual definition.
We note that $B(\rho,\rho')$ vanishes if and ony if the states have orthogonal supports, $\rho\rho'=0$, providing a measure of distinguishability. Although it produces less tight bounds on the probability of discrimination error than the  quantum Chernoff bound \cite{Acin:2008, Chernoff:1952}, it has the property of factorizing w.r.t. the tensor product, making its calculation easier and similar to the decoherence factor:
 \begin{align}
B^{\text{mac}}_{X_0,X'_0}\equiv B[\rho_{\text{mac}}(X_0), \rho_{\text{mac}}(X'_0)]=\prod_{k \in \text{mac}} B^{(k)}_{X_0,X'_0}. \label{Bmac}
\end{align}
From the metrological point of view, encoding $X_0$ into a completely separable state (\ref{mac}) leads to a rather uninteresting classical scenario \cite{Braunstein:1994zz}, but encoding efficiency is not our goal here and we postpone a study of more advanced scenarios with entangled macrofraction states to a future research. Just like with the decoherence, we expect that distinguishability will be achieved only at some characteristic length scale.

Summarizing, the approach to the SBS structure will be characterized by two quantities \cite{Mironowicz:2017}, the decoherence factor (\ref{Gamma}) and the generalized overlap (\ref{Bmac}). We will call them ``objectivity markers''.
We note that their factorized character, i.e. a total measure is a product of measures for individual environmental systems, which is due to the uncorrelated initial state. As one expects, a single factor corresponding to a single environmental spin, will be oscillatory and of course will not lead anywhere close to the SBS structure. However due to the factorized character, we expect that for a sufficiently large groups of spins a considerable dephasing will take place at some length scales, leading to an approximate SBS structure. 

\section{Calculation of objectivity markers}
Here without loss of generality we consider only a single spin environment in each expression (\ref{Gamma}, \ref{Bmac}), dropping the environment indices. 
State fidelity has a particularly simple form for spin-$\frac{1}{2}$ states. Let $M\equiv\sqrt{\rho}\rho'\sqrt{\rho}$, then:
\begin{align}
   B(\rho,\rho')
   &=\text{Tr}M+2\sqrt{\det M}.\label{B2}
\end{align}
Since in our case $\rho$ and $\rho'$ have the same initial state $\rho_0$ but different evolutions, $U$ and $U'$,  we obtain:
\begin{align*}
    M=U\sqrt{\rho_0}U^\dagger U'\rho_0 U'^\dagger U\sqrt{\rho_0}U^\dagger,
\end{align*}
and finally:
\begin{align*}
   B_{X_0,X'_0} =\text{Tr}\left[U^\dagger_{X_0,X'_0}\rho_0U_{X_0,X'_0}\rho_0\right]+2\det\rho_0,
\end{align*}
where we defined a relative evolution operator:
\begin{align}
   U_{X_0,X'_0}\equiv U^\dagger(X'_0,t)U(X_0,t).\label{relative_U}
\end{align}
We note that the decoherence factor also depends on it. 

Before calculating the markers for the evolution (\ref{Uk}) , we first derive general expressions.
We will find it convenient to use the Bloch representation, decomposing any operator into identity and Pauli matrices. This representation will give a nice geometrical interpretation for the decoherence factor and the generalized overlap and their complementarity relation. Let us decompose the initial state and the relative evolution in the Pauli basis:
\begin{align}
    \rho_0=\frac{1}{2}\left(1+\vec{a}\cdot\vec{\sigma}\right),
    \label{rho_0}
\end{align}
where  $|\vec{a}|\leq1$ and
\begin{align}
U_{X_0,X'_0}=u_0+i\vec{u}\cdot\vec{\sigma},\label{U_total_Pauli}
\end{align}
where  
\be
u^2_0+|\vec{u}|^2=1.\label{u2}
\ee 
Then it is easy to obtain the decoherence factor:
\begin{align}
\Gamma_{X_0,X'_0}=u_0+i\vec{a}\cdot \vec{u}\label{Gamma_u_a}
\end{align}
and its modulus:
 \begin{align}
|\Gamma_{X_0,X'_0}|^2&=u^2_0+(\vec{a}\cdot\vec{u})^2,\label{decoherence factor magintude}
\end{align}
which controls the decoherence process in the position basis. Calculation of the generalized overlap in turn, leads to
(see Appendix ~\ref{AppA} for the details):
\begin{align}
    B_{X_0,X'_0}=1-|\vec{a}\times \vec{u}|^2.\label{B^2_a_u}
\end{align}
Combining \eqref{decoherence factor magintude} and \eqref{B^2_a_u} the relation between $|\Gamma_{X_0,X'_0}|^2$ and $B_{X_0,X'_0}$ is expressed by:
\begin{align}
    B_{X_0,X'_0}-|\Gamma_{X_0,X'_0}|^2=(1-|\vec{a}|^2)(1-u^2_0).\label{B^2_Gamma^2}
\end{align}
This relation can be interpreted as a sort of complementarity between decoherence and distinguishability.

Using \eqref{decoherence factor magintude} and \eqref{B^2_Gamma^2}, we can express the decoherence factor and the generalized overlap in the high-frequency expansion of Sec. \ref{Dynamics}. As seen, the decoherence factor and the generalized overlap are a function of only a vector $\vec{a}$ representing an initial density matrix for a spin $\rho_0$ and a relative unitary operator  $U_{X_0,X'_0}$ defined in \eqref{relative_U}.  Using (\ref{Uk}), we obtain in the lowest non-zero order of the high-frequency expansion
\begin{align}\nonumber
 U_{X_0,X'_0}&= U^\dagger_{F}(\tau;\xi') U_K(\tau;\xi-\xi')U_{F}(\tau;\xi)\\
&=u_0+i\vec{u}\cdot\vec{\sigma},\label{relative_U_Floquet}
\end{align}
where 
\begin{align}\nonumber
 u_0&=\cos[\tilde{\Delta}(\xi^2-\xi'^2)\tau]\cos[\delta\xi\sin\tau],\\\nonumber
 u_1&=-\sin[\tilde{\Delta}(\xi^2-\xi'^2)\tau]\cos[\delta\xi\sin\tau],\\\nonumber
 u_2&=\sin[\tilde{\Delta}(2-\xi^2-\xi'^2)\tau]\sin[\delta\xi\sin\tau],\\
 u_3&=-\cos[\tilde{\Delta}(2-\xi^2-\xi'^2)\tau]\sin[\delta\xi\sin\tau],\label{ui_High_Frequency}
\end{align}
with the notation $\delta\xi\equiv\xi-\xi'$ and the dimensionless parameters defined in (\ref{parameters}). We recall that 
for the trajectories (\ref{traj}) there is no initial kick at $t_0=0$ [cf. (\ref{K})].
Each $u_i$ in (\ref{ui_High_Frequency}) is a product of a fast and a slow moving parts, as one would expect from the Floquet theory. The frequency of the slow motion is proportional to the tunneling energy $\tilde{\Delta}$, while the fast-moving terms are independent of it and have the time-dependent frequency $\delta\xi\sin\tau$. There is also a distinction between $(u_0,u_1)$ and $(u_2,u_3)$. The $(u_0,u_1)$ pair has large overall sinusoidal patterns with small internal vibrations while in $(u_2,u_3)$ the overall profiles are comparable to internal vibrations. As we will see, the fast moving parts are unwanted for objectivity.

\begin{figure}[t!]
\includegraphics[width=0.9\linewidth, height=4cm]{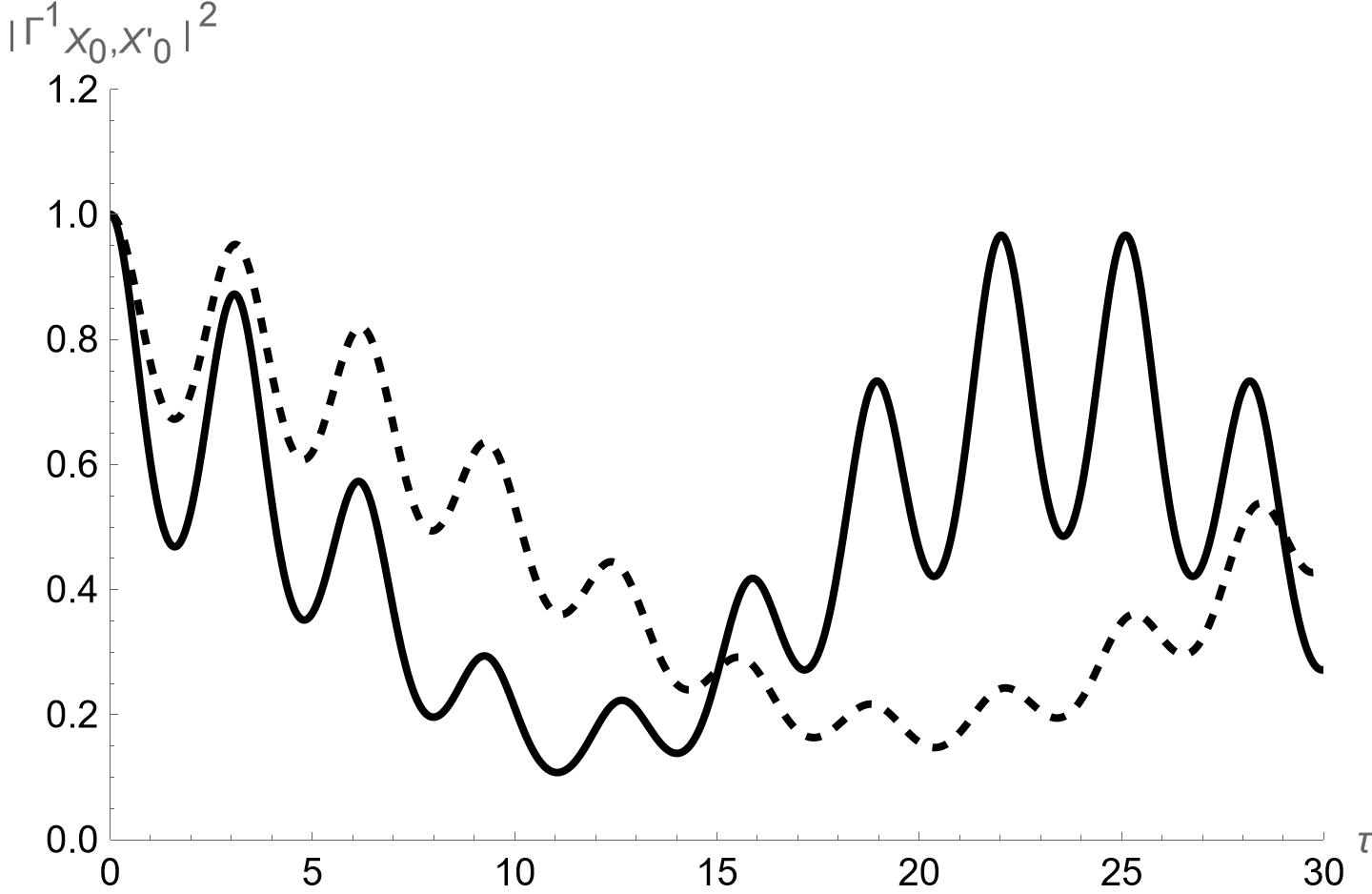} 
\caption{\label{fig:Gamma1}Time dependence of decoherence factors for a single spin environment \eqref{G1}. The solid and dashed lines stand for $\xi=0.9$ and $0.6$, respectively, while $\xi'=0.1$ is fixed. The rest of the parameters are $\tilde{\Delta}=\frac{1}{6}, \beta\Delta=1$. Both slow and fast oscillations are clearly visible.}
\end{figure}

\begin{figure}[t!]
\includegraphics[width=0.9\linewidth, height=4cm]{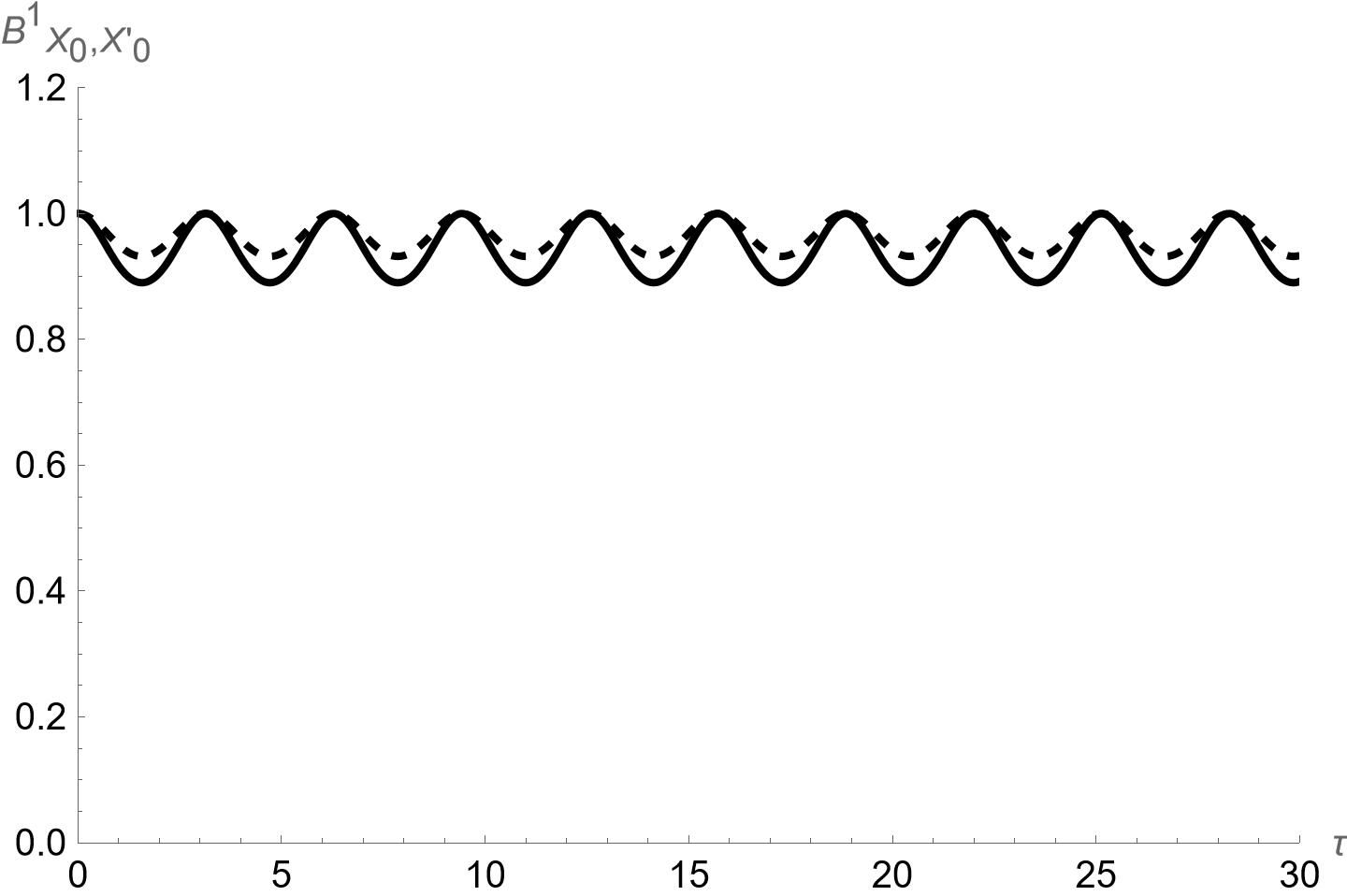} 
\caption{Time dependence of generalized overlaps for a single spin environment \eqref{B1}. The solid and dashed lines stand for $\xi=0.9$ and $0.6$, respectively, while $\xi'=0.1$ is fixed. The rest of the parameters are the same as in Fig.~\ref{fig:Gamma1} for a better comparison: $\tilde{\Delta}=\frac{1}{6}, \beta\Delta=1$.} 
\label{fig:B1}
\end{figure}

To proceed further, we will assume  that the environment is initially in the thermal state, i.e.
\begin{align}\label{rthm}
    \rho^{(k)}_E(0)=\frac{e^{-\beta H_E^{(k)}}}{\text{Tr}[e^{-\beta H_E^{(k)}}]}=\frac{1}{2}\left[1+\sigma_x\tanh\left(\frac{\beta\Delta_k}{2}\right)\right],
\end{align}
where $\beta=1/k_BT$, so that the parameters of the initial state are given by  $\vec{a}=(E(\beta),0,0)$, where we introduced:
\begin{align}\label{EB}
    E(\beta)\equiv\tanh\left(\frac{\beta\Delta}{2}\right)=\frac{\langle E\rangle}{\Delta/2},
\end{align}
 denoting the average thermal energy, rescaled by the tunneling energy.
 We  obtain the following single-factor expressions for the decoherence and the generalized overlap:
\begin{align}
    |\Gamma^1_{X_0,X'_0}|^2&=u^2_0+E(\beta)^2u^2_1\\
    B^1_{X_0,X'_0}&=1-E(\beta)^2+E(\beta)^2(u^2_0+u^2_1).\label{Gamma_B_ax}
\end{align}
which leads to:
 \begin{align}\label{G1}
&|\Gamma^1_{X_0,X'_0}|^2 =\left[1-\frac{\sin^2[\tilde{\Delta}(\xi^2-\xi'^2)\tau]}{\cosh^2(\beta\Delta/2)}\right]\cos^2[\delta\xi\sin\tau]\\
 &B^1_{X_0,X'_0}=1-E(\beta)^2\sin^2[\delta\xi\sin\tau].\label{B1}
\end{align}
The time dependence of the decoherence  factor is given by the slow-motion and  the fast-motion modulating each other. In contrast,  the generalized overlap depends only on the fast oscillating part. 
We note that the decoherence factor depends here on the temperature, contrasting the result in a boson-spin system mapped from that of the quantum Brownian motion in the Born-Markov approximation. \cite{Schlosshauer2007}. In particular, at zero temperature the slow-motion part vanishes. 
Sample plots of the markers as the functions of the rescaled time $\tau$ are presented in Figs.~\ref{fig:Gamma1} and \ref{fig:B1}.

The full decoherence and overlap functions are products of the above factors [cf. \eqref{Gamma} and \eqref{Bmac}]. 
We begin their analysis by first assuming small dimensionless amplitudes of the central oscillator $\xi, \xi' \ll 1$, which allows to expand the trigonometric functions. In particular,  
$\sin^2[\tilde{\Delta}(\xi^2-\xi'^2)\tau] \approx (\xi^2-\xi'^2)^2 t^2\Delta^2/4 + O(\xi^8)$, which is valid for times:
\be
t \ll \frac {2\Omega^2}{g^2\Delta|X_0^2-X^{'2}_0|}.
\ee
Similarly, we expand the fast motion factors containing $[(\xi-\xi')\sin\tau]$. Keeping the terms at most quadratic in $\xi$, we obtain:
\begin{align}\nonumber
|\Gamma^1_{X_0,X'_0}|^2& =1-\delta\xi^2\sin^2\tau +O(\xi^4) \\
&\approx \exp\left[-\frac{g^2 \delta X_0^2}{\Omega^2} \sin^2\Omega t\right],\label{Gamma_2nd}
\end{align}
where we returned to the original variables having defined (\ref{parameters}) and defined $\delta X_0 \equiv X_0-X_0'$. The temperature dependence, the whole slow oscillating part, disappears in the lowest order and appears only in the $\xi^4$ terms and higher. The behavior of the generalized overlap is different in this respect and depends strongly on the temperature even in the lowest order in $\xi$:
\begin{align}\nonumber
B_{X_0,X'_0}^1& =1-\delta\xi^2E(\beta)^2\sin^2\tau +O(\xi^4) \\
&\approx \exp\left[-\frac{g^2 \delta X_0^2}{\Omega^2}E(\beta)^2\sin^2\Omega t\right].\label{B_2nd}
\end{align}
To calculate the products \eqref{Gamma} and \eqref{Bmac}, we will assume that the constants $g_k$ and $\Delta_k$ are identically and independently distributed according to some probability distributions  and there are sufficiently many terms in the resulting sums in the exponentials to apply  the law of large numbers:
\begin{align}
|\Gamma_{X_0,X'_0}|^2& =\prod_{k \in uE}|\Gamma^{(k)}_{X_0,X'_0}|^2 = \exp\left[-\frac{\delta X_0^2}{\Omega^2} \sum_{k\in uE}g_k^2\sin^2\Omega t\right]\\
& \approx \exp\left[-N_u\frac{\langle g^2\rangle \delta X_0^2}{\Omega^2} \sin^2\Omega t\right],\label{Gapprox}
\end{align}
where $N_u$, is the size of the unobserved fraction $uE$ of the environment and $\langle g^2\rangle$ is the average of $g_k$ over the $uE$. 
This procedure \cite{Mironowicz:2017} can be viewed as a form of an introduction of spectral density, but we keep control of the number of spins in the fractions.
Similarly for the generalized overlap:
\begin{align}
B_{X_0,X'_0}& =\prod_{k \in \text{mac}} B^{(k)}_{X_0,X'_0}\nonumber\\
&\approx \exp\left[-N_{\text{mac}}\frac{\langle g^2\rangle\delta X_0^2}{\Omega^2} \left \langle E(\beta)^2\right\rangle  \sin^2\Omega t\right],\label{Bapprox}
\end{align}
where $N_{\text{mac}}$ is the observed macrofraction size and we assumed that the distributions of $g$s' and $\Delta$s' are independent; $\left \langle E(\beta)^2\right\rangle $ is understood as the average over the $\Delta$ [cf. \eqref{EB}].
Due to the periodicity of the markers \eqref{Gapprox} and \eqref{Bapprox}, it is immediately obvious that in the small displacement limit, there are complete recoherences at the turning points $t_n=n\pi/\Omega$ and there is no asymptotic behavior as $t\to\infty$. We can thus speak of 
the approach to the objective state only in the time intervals between the turning points. As anticipated, this approach is governed by two length scales, controlling the decoherence and the distinguishability processes:
\begin{align}
\lambda_{\text{dec}} &\equiv \frac{\Omega}{\sqrt{N_u\langle g^2\rangle}},\label{ldec}\\
\lambda_{\text{dist}}& \equiv \frac{\Omega} {\sqrt{N_{\text{mac}}\langle g^2\rangle \left \langle E(\beta)^2\right\rangle}}.\label{ldist}
\end{align}
Their dependencies on the fraction sizes, or equivalently on the Hilbert space dimensionalities of the fractions, mean that shorter distances are resolved as more spins are taken into account. We call the above length scales the decoherence and the distinguishability length scales. The distinguishability length scale is temperature-dependent and grows with the temperature approximately linearly for high temperatures. This is intuitively clear as hotter environment is closer to the totally mixed state and thus its information-carrying capabilities are worse. Moreover, for non-zero temperatures, $ \lambda_{\text{dist}} > \lambda_{\text{dec}}$ for the same fraction sizes, meaning that one can extract  the position $X_0$ from the environment with a worse resolution than one at which the environment decoherences the central system. This phenomenon of ``bound information'' in the environment was observed in the QBM model in \cite{Lee:2023qrm}.

\begin{figure}[t!]
\includegraphics[width=0.9\linewidth, height=4cm]{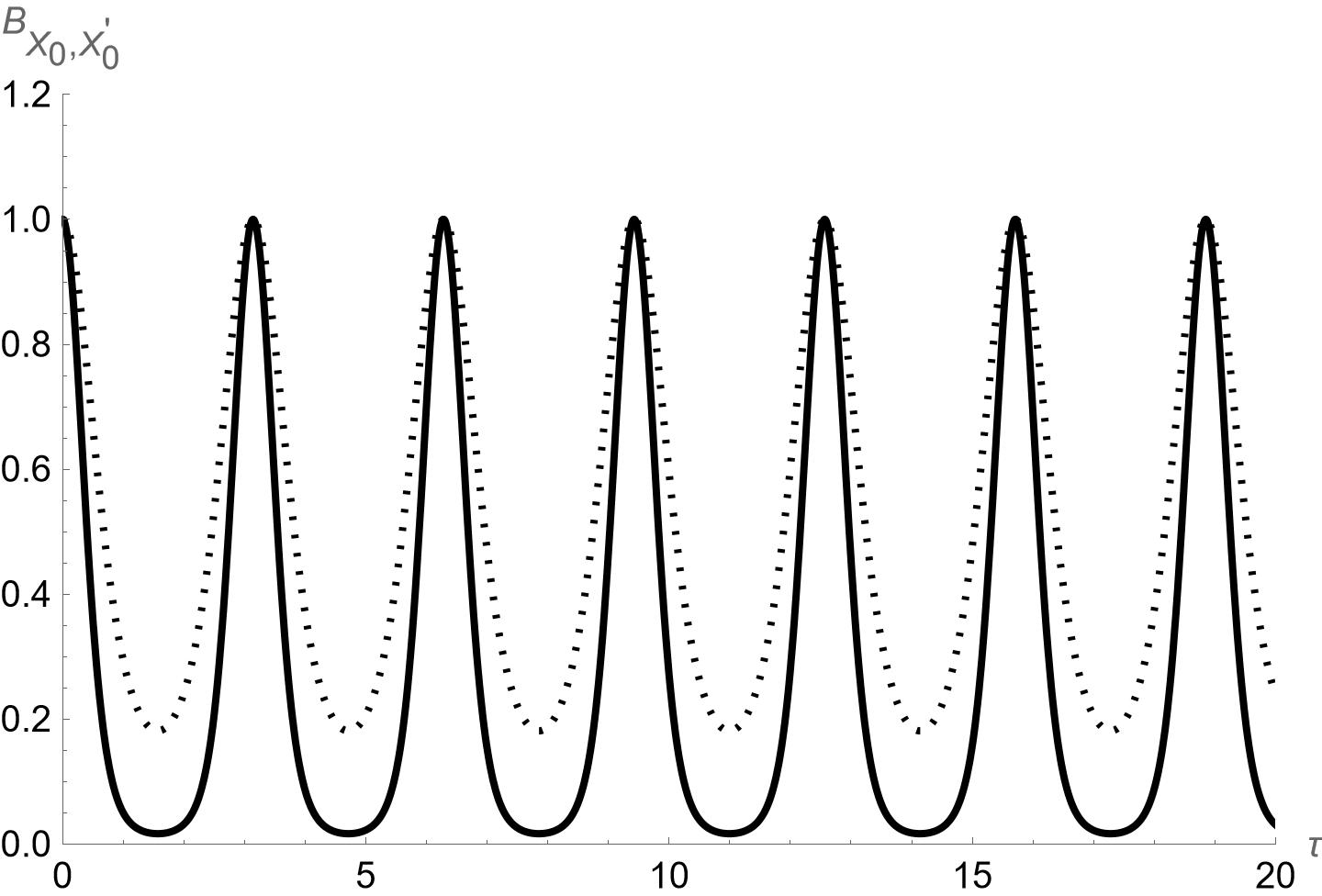} 
\caption{Generalized overlap for $N_{\text{mac}}=100$ spin environments with different interaction couplings $\xi$ (the solid and dashed lines stand for $\bar\xi=0.9$ and $0.6$, respectively.). $(\tilde{\Delta}=\frac{1}{6},\Omega=3,\beta\Delta=1,\bar\xi'=0.1)$ are chosen.}
\label{fig:B}
\end{figure}

\begin{figure}[t!]
\includegraphics[width=0.9\linewidth, height=4cm]{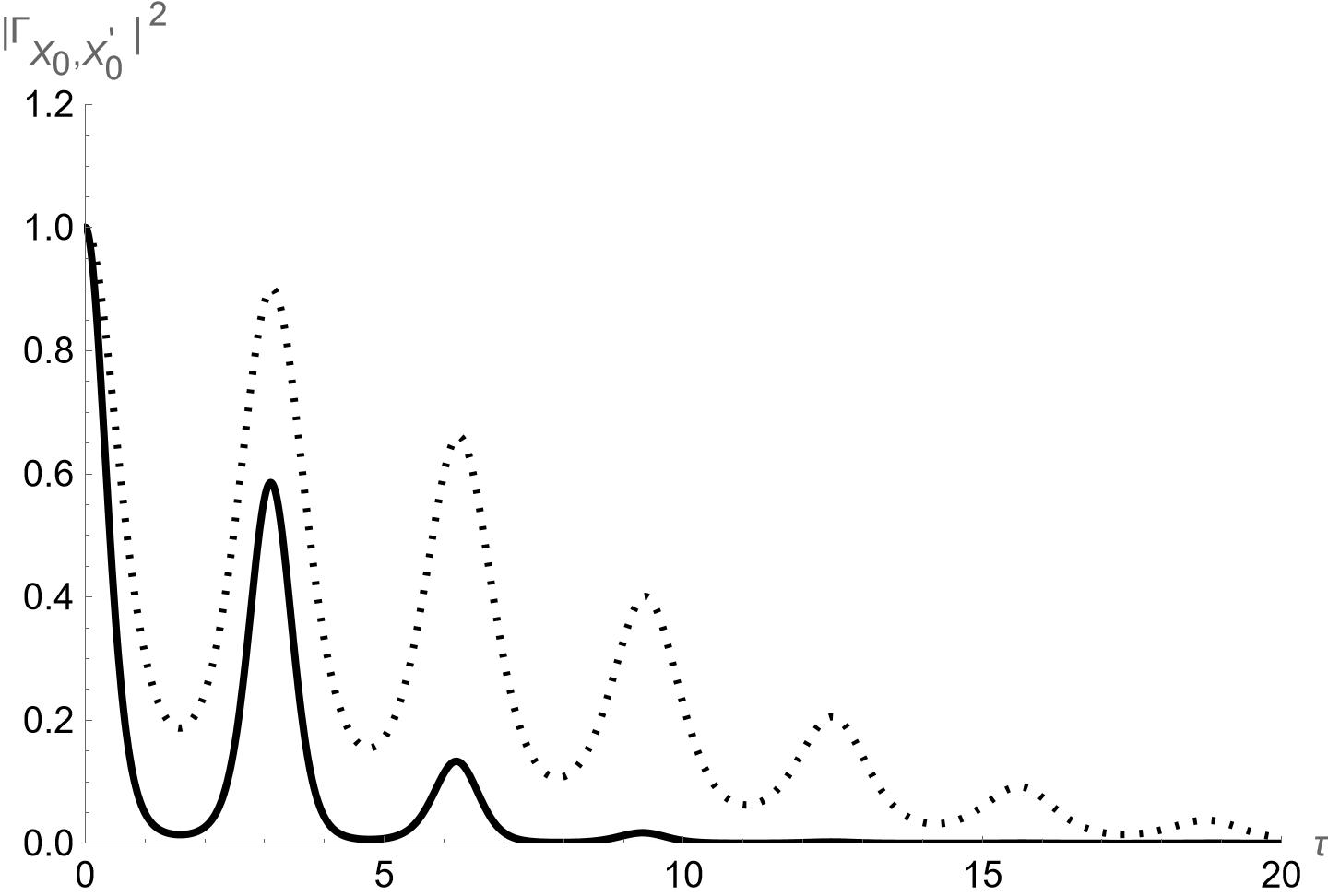} 
\caption{Decoherence factors for $N_u=20$ spin environments with different interaction couplings $\xi$ (the solid and dashed lines stand for $\xi=0.9$ and $0.6$, respectively.). $(\tilde{\Delta}=\frac{1}{6},\Omega=3,\beta\Delta=1,\xi'=0.1)$ are chosen.}
\label{fig:Gamma}
\end{figure}

Let us now analyze the objectivity markers beyond the small-amplitude approximation. We first look at the generalized overlap:
\begin{align}
    B_{X_0,X'_0}=\prod_{k\in mac}\left[1-E_k(\beta)^2\sin^2(\delta\xi_k\sin\tau)\right].
\end{align}
 It is immediately clear that the time-periodic frequency $\sin\tau$ dictates the periodic character of $B_{X_0,X'_0}$:
\be
 B_{X_0,X'_0}(t)=B_{X_0,X'_0}(t+\pi/\Omega),
\ee
irrespectively of what are the distributions of $g_k$ and $\Delta_k$. In particular, this periodicity implies a complete loss of distinguishability, $B_{X_0,X'_0}(t_n)=1$, at the turning points $t_n=n\pi/\Omega$, just like in the approximate analysis above.
 The behavior of $B_{X_0,X'_0}(t)$ can be approximated using the Law of Large Numbers in the following way:
\begin{align}\nonumber
    \log B_{X_0,X'_0} & =\sum_{k\in \text{mac}} \log  B^{(k)}_{X_0,X'_0} \approx N_{\text{mac}} \langle \log B_{X_0,X'_0}^1\rangle \\
    & \leq N_{\text{mac}}  \log \langle B_{X_0,X'_0}^1\rangle,\label{LLN1}
\end{align}
where we used the concavity of the logarithm. Furthermore:
\begin{align}\nonumber
 \langle B_{X_0,X'_0}^1\rangle = 1-\left \langle E(\beta)^2\right\rangle\left\langle\sin^2(\delta\xi\sin\tau)\right\rangle.
\end{align} 
We are interested in the last average as it determines the time dependence. For simplicity we will assume a uniform distribution of $g$ over some interval $[0, \bar g]$. This corresponds to a spectral density with a sharp cut-off at $\bar g$.
Elementary integration gives:
\begin{align}
& \left\langle\sin^2(\delta\xi\sin\tau)\right\rangle  = \frac{1}{\bar g}\int_0^{\bar g} dg \sin^2\left[\frac{g\delta X_0}{\Omega}\sin\tau\right] \label{B1avg1}\\
& = \frac{1}{2}\left[1-\sinc \left(\frac{2\bar g \delta X_0}{\Omega}\sin\tau\right)\right],\label{B1avg2}
\end{align}
where $\sinc x\equiv \sin x/x$, which leads to:
\begin{align}
 & B_{X_0,X'_0}  \approx \nonumber\\
 & \left[ 1-\frac{1}{2}\left \langle E(\beta)^2\right\rangle \left(1-\sinc \left(\frac{2\bar g \delta X_0}{\Omega}\sin\tau\right)\right)\right]^{N_{\text{mac}}}. \label{Bfinal}
\end{align}
Since $1-\sinc x \approx x^2/6  +O(x^4)$, rises(decays) around the turning points $t_n=n\pi/\Omega$ are given by the small amplitude approximation (\ref{Bapprox}), with $\lambda_{dist}$ rescaled by an unimportant factor $\sqrt 3$.
A sample plot of (\ref{Bfinal}) is presented in  Fig. ~\ref{fig:B} for different values of $\delta\bar\xi=\bar g \delta X_0/\Omega$.
The decoherence factor can be analyzed in the same steps \eqref{LLN1}-\eqref{Bfinal} with the complication that it is composed of both slow-and fast-moving parts. We need the average of (\ref{G1}). For simplicity we will assume $\Delta_k = \Delta$ so the only randomness is in $g_k$:
 \begin{align}\nonumber
&\left\langle|\Gamma_{X_0,X'_0}|^2\right\rangle= \left\langle\cos^2(\delta\xi\sin\tau)\right\rangle \\\nonumber
& -\frac{\left\langle\sin^2[\tilde{\Delta}(\xi^2-\xi'^2)\tau]\cos^2(\delta\xi\sin\tau)\right\rangle}{\cosh^2(\beta\Delta/2)}\label{fast}\\
&=\frac{\cosh(\beta\Delta)}{\cosh(\beta\Delta)+1}\left\langle\cos^2(\delta\xi\sin\tau)\right\rangle\\
&+\frac{1}{2\cosh(\beta\Delta)+2}\bigg\{\left\langle\cos[2\tilde{\Delta}(\xi^2-\xi'^2)\tau]\right\rangle\label{slow1}\\
&+\left\langle\cos[2\tilde{\Delta}(\xi^2-\xi'^2)\tau]\cos(2\delta\xi\sin\tau)\right\rangle\bigg\},\label{slow2}\\
& \equiv \Gamma_{\text{fast}}(\tau) + \Gamma_{\text{slow}}(\tau),
\end{align}
where we used trigonometric and hyperbolic identities to simplify the expressions. The first term \ref{fast} is the fast oscillating part which is equal to: 
\begin{align}
&\Gamma_{\text{fast}}(\tau)\nonumber\\
&=\frac{\cosh(\beta\Delta)}{2\cosh(\beta\Delta)+2}\left[1 +\sinc \left(\frac{2\bar g \delta X_0}{\Omega}\sin\tau\right)\right],\label{Gfast}
\end{align} 
where we used above the same averaging as in \eqref{B1avg1} and \eqref{B1avg2}. It is clearly time-periodic due to the periodic frequency $\sin\tau$, just like \eqref{Bfinal}, but it is multiplied by a temperature-dependent factor that is always smaller than 1.  The behavior around the turning points $t_n=n\pi/\Omega$ is again given by the Gaussian law (\ref{Gapprox}) with the $\lambda_{\text{dec}}$ rescaled by $\sqrt 3$.
The terms \eqref{slow1} and \eqref{slow2} are the slow oscillating parts, contributing for non-zero temperatures. 
They can be calculated explicitly for a uniform distribution by using Fresnel integrals as they
contain $g^2$ under the cosine. For example, the term \eqref{slow1} is proportional to:
\begin{align}
& f(\tau)\equiv\left\langle \cos[2\tilde{\Delta}(\xi^2-\xi'^2)\tau]\right\rangle \nonumber\\
& =\frac{1}{\sqrt{2\tilde\Delta(\bar\xi^2-\bar\xi'^2)\tau}} C\left(\sqrt{2\tilde\Delta(\bar\xi^2-\bar\xi'^2)\tau}\right),\label{C}
\end{align}
where $C(x)\equiv \int_0^x du \cos u^2$ is the cosine Fresnel integral and $\bar\xi\equiv \bar g X_0/\Omega$. The long-time behavior of this term is determined by an asymptotic expansion for large $x$,  
$C(x)\approx \sqrt{\pi/8} + \sin x^2/2x +O(x^{-3})$ \cite{Gradshtein:1980}, which gives $f(\tau) \sim 1/\sqrt\tau$ for long times.
The last term, \eqref{slow2}, is a bit more complicated but can be manipulated using basic trigonometric identities (Refer to Appendix (\ref{A:slow4}).
As a result, from \eqref{C} and \eqref{A:slow4} it follows that for large times $\Gamma_{\text{slow}}(\tau) \sim 1/\sqrt\tau$. And finally:
\begin{align}
&|\Gamma_{X_0,X'_0}|^2 \approx \left[ \Gamma_{fast}(\tau)+\Gamma_{slow}(\tau)\right]^{N_u} \label{Gfinal}\\
&= [\Gamma_{fast}(\tau)]^{N_u} + O\left(\frac{1}{\sqrt\tau}\right)\\
& =\left[\frac{\cosh(\beta\Delta)}{\cosh(\beta\Delta)+1}\right]^{N_u}\left[\frac{1}{2} +\frac{1}{2}\sinc \left(\frac{2\bar g \delta X_0}{\Omega}\sin\tau\right)\right]^{N_u}\label{Gfinal2}\\
&  + O\left(\frac{1}{\sqrt\tau}\right)\nonumber
 \end{align}
Despite the presence of a time-periodic term, unlike the generalized overlap \eqref{Bfinal} this function can effectively 
decay with time, meaning decoherence can take place.  This is due to the temperature-dependent prefactor in \eqref{Gfinal2}, which 
multiplies the $\sinc$-term and which decays with temperature and for high temperatures (small $\beta$) is of the order $\sim 2^{-N_u}$. Thus, although the $\sinc$-term is equal to one at the turning points $t_n=n\pi/\Omega$, its contribution is damped by the temperature-dependent term.  A sample plot of the exact expression (\ref{Gfinal}) using \eqref{Gfast},\eqref{C} and \eqref{A:slow4} is presented in  Fig. ~\ref{fig:Gamma}.

We conclude that for cosine trajectories \eqref{traj}, although a central oscillator can be effectively decohered, the environment however is unable to reliably store the amplitude information for all times as there are  
periodic and complete losses of distinguishability at the turning points. Thus,  objective states can form only between the turning points. 
We suspect these perfect revivals are caused by the recoil-less approximation, which completely neglects the damping of the central oscillator. They are also in contrast with the oscillator environment, where for the same trajectory \eqref{traj} and under the same approximation (\ref{final}) no such revivals were observed, but rather a steady decay \cite{Tuziemski:2015}. The revivals, in turn, appeared in the QBM model for sine trajectories, corresponding to initial position squeezing, which was in agreement with the earlier works \cite{Paz:1992}. It is therefore interesting to study more general trajectories in the current model too.

\subsection{Arbitrary Trajectory}
\begin{figure}[t!]
\includegraphics[width=0.9\linewidth, height=4cm]{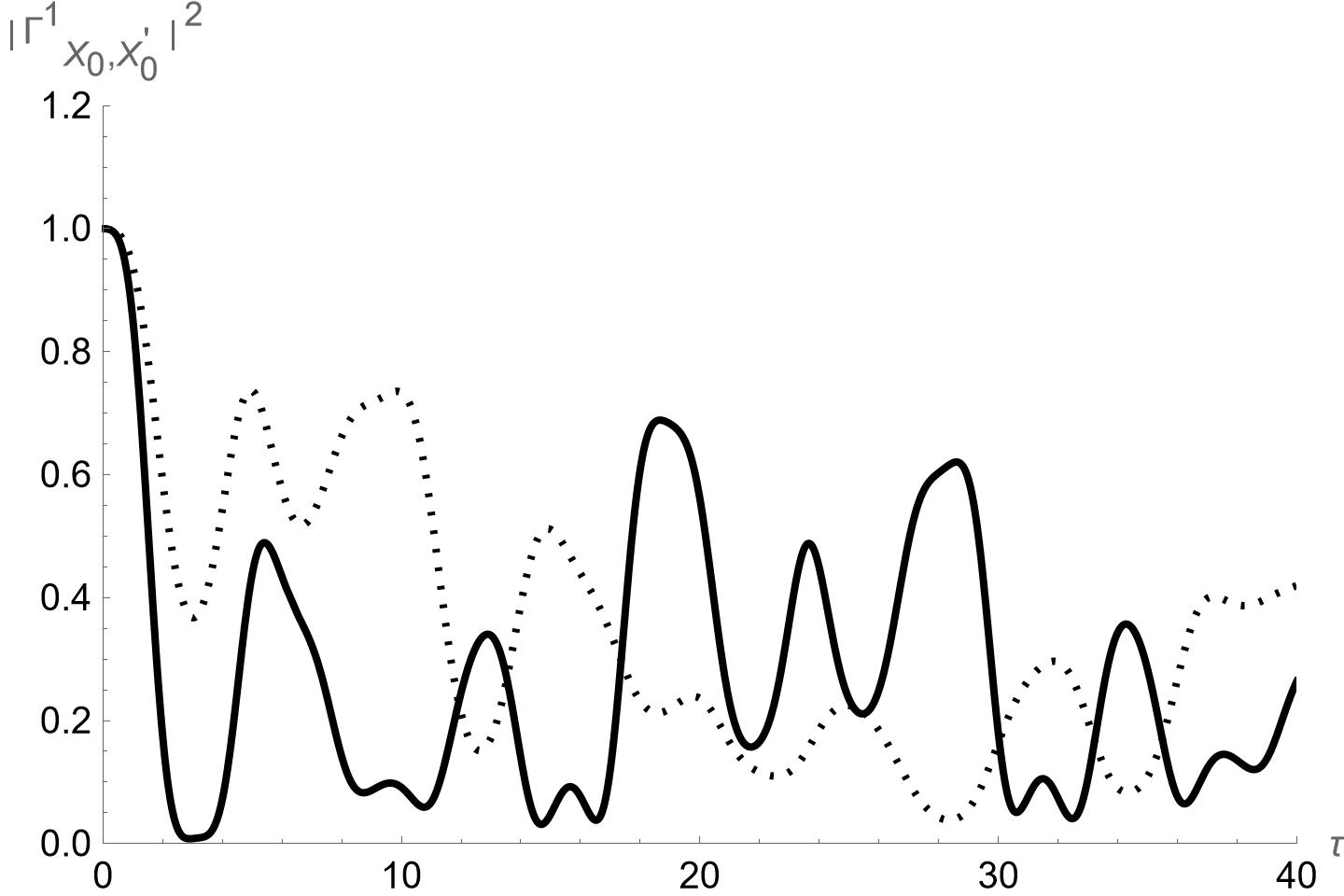} 
\caption{Time dependence of decoherence factor for a single spin environment with $\phi=\pm\frac{\pi}{2}$. The solid and dashed lines correspond to $\xi=0.9$ and $0.6$, respectively, while $\xi'=0.1$ is fixed. The rest of the parameters are $\tilde{\Delta}=\frac{1}{6},\beta\Delta=1$. The separation between a slow and fast oscillations is less clear due to involvement of another frequency than for $\phi=0$ in Fig.\ref{fig:Gamma1}}
\label{fig:Gamma1_0.5pi}
\end{figure}

\begin{figure}[t!]
\includegraphics[width=0.9\linewidth, height=4cm]{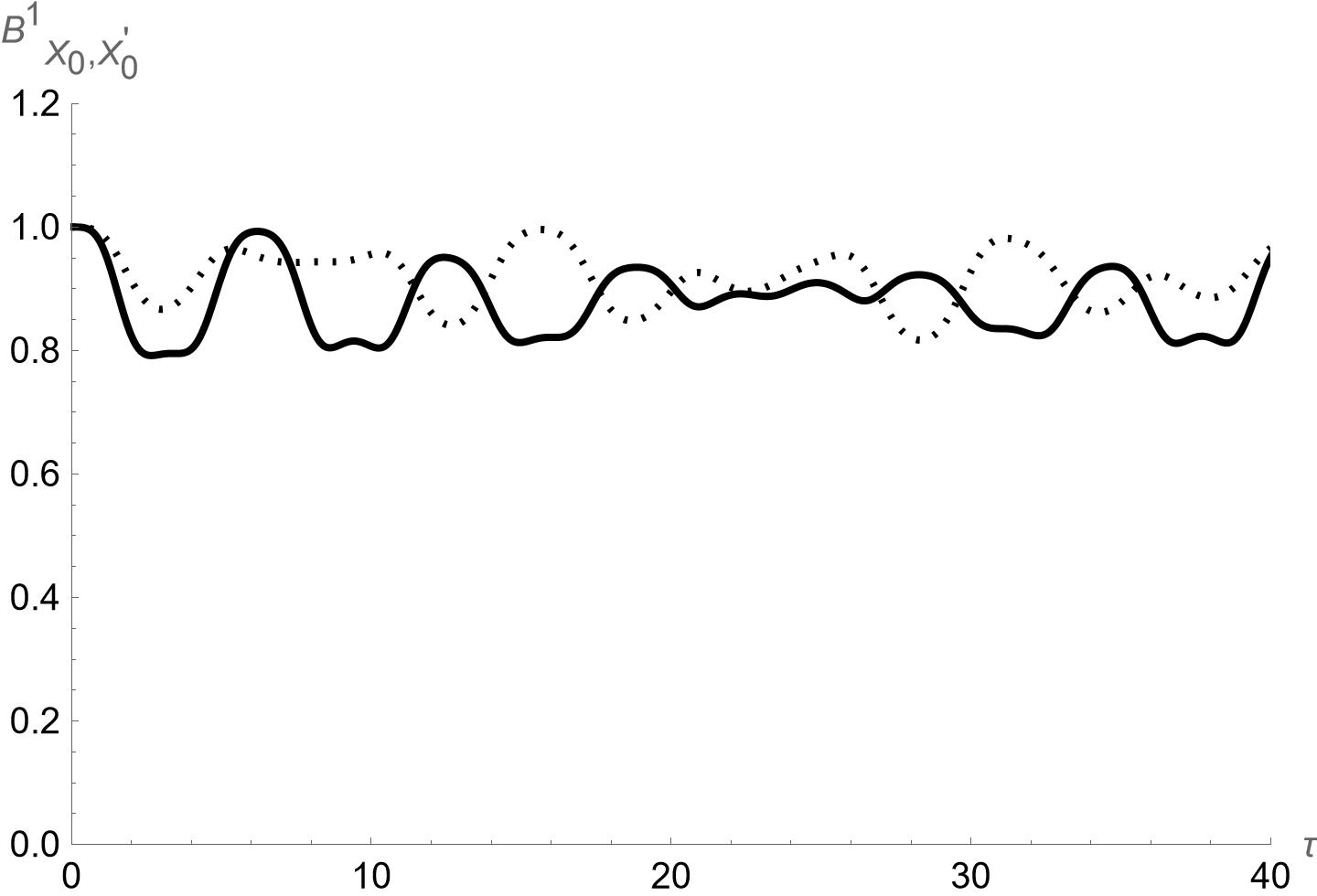}
\caption{Time dependence of the generalized overlap for a single spin environment with $\phi=\pm\frac{\pi}{2}$. The solid and dashed lines correspond to $\xi=0.9$ and $0.6$, respectively, while $\xi'=0.1$ is fixed. The rest of the parameters are $\tilde{\Delta}=\frac{1}{6},\beta\Delta=1$. The separation between a slow and fast oscillations is less clear due to involvement of another frequency than for $\phi=0$ in Fig.\ref{fig:B1}}
\label{fig:B1_0.5pi}
\end{figure}

An arbitrary trajectory of the central oscillator is obtained by adding a constant phase $\phi$ to (\ref{traj}):
\begin{align}\label{traj2}
    X(t)=X_0\cos(\tau+\phi).
\end{align}
It is then interesting  to investigate how this phase can affect the objectivity, especially whether there is a possibility to overcome the asymmetry between the decaying decoherence factor and the monotonously oscillating generalized overlap 
found above. The phase changes the Fourier components $V^{(1)}$ and $V^{(-1)}$ in the high-frequency expansion in (\ref{Hi}),
\begin{align}
    V^{(\pm1)}=\frac{gX_0}{2}\sigma_z\to\frac{gX_0}{2}\sigma_ze^{\pm i\phi}.
\end{align}
Consequently, as seen in (\ref{def:He}) and (\ref{def:K}), $K(t)$ gets a phase change (\ref{K}): 
\begin{align}
      K(t)&=\xi\sigma_z\sin(\tau+\phi),
\end{align}
while the Floquet Hamiltonian $H_F$ remains the same as in (\ref{He}).
In (\ref{Floquet_U}) $\phi\neq0$ contributes to $U_{X_0,X'_0}$ due to the non-trivial initial kicks $K(0)$. Explicitly, from  (\ref{Floquet_U}) and (\ref{relative_U_Floquet}), 
we obtain $U_{X_0,X'_0}=u_0+i\vec{u}\cdot\vec{\sigma}$ with [cf. \eqref{ui_High_Frequency}]
\begin{align}\label{ui_phi}\nonumber
    u_0&=\cos[\delta \xi\sin\phi]\cos[\tilde{\Delta}(\xi^2-\xi'^2)\tau]\cos[\delta \xi\sin(\tau+\phi)]\\\nonumber
    &+\sin[\delta \xi\sin\phi]\cos[\tilde{\Delta}(\xi^2+\xi'^2-2)\tau]\sin[\delta \xi\sin(\tau+\phi)]\\\nonumber
    u_1&=-\cos[(\xi+\xi')\sin\phi]\sin[\tilde{\Delta}(\xi^2-\xi'^2)\tau]
    \\\nonumber
    &\times\cos[\delta \xi\sin(\tau+\phi)]\\\nonumber
    &+\sin[(\xi+\xi')\sin\phi]\sin[\tilde{\Delta}(\xi^2+\xi'^2-2)\tau]\\\nonumber
    &\times\sin[\delta \xi\sin(\tau+\phi)]\\\nonumber
     u_2&=-\sin[(\xi+\xi')\sin\phi]\sin[\tilde{\Delta}(\xi^2-\xi'^2)\tau]\\\nonumber
    &\times\cos[\delta \xi\sin(\tau+\phi)]\\\nonumber
&-\cos[(\xi+\xi')\sin\phi]\sin[\tilde{\Delta}(\xi^2+\xi'^2-2)\tau]\\
    &\times\sin[\delta \xi\sin(\tau+\phi)]\\\nonumber
     u_3&=
    \sin[\delta \xi\sin\phi]\cos[\tilde{\Delta}(\xi^2-\xi'^2)\tau]\cos[\delta \xi\sin(\tau+\phi)]\\\nonumber
    &-\cos[\delta \xi\sin\phi]\cos[\tilde{\Delta}(\xi^2+\xi'^2-2)\tau]\sin[\delta \xi\sin(\tau+\phi)].
\end{align}
The only relevant components for the decoherence factor \eqref{decoherence factor magintude} and the generalized overlap \eqref{B^2_a_u} associated with the thermal state \eqref{rthm}, i.e. with $\vec{a}=(E(\beta),0,0)$, are $u_0$ and $u_1$, 
which follows from \eqref{u2}. We first analyze the small $\xi$ approximation as it is easier. The factors $u^2_0$ and $u^2_1$ then read as
\begin{align}\nonumber
    u^2_0&=1-\delta \xi^2[\sin^2\phi+\sin^2(\tau+\phi)\\
&-2\sin\phi\sin(\tau+\phi)\cos(2\tilde{\Delta}\tau)]+O(\xi^4),\\\nonumber
    u^2_1&=O(\xi^4),\label{u0_u1_phi}
\end{align}
which from \eqref{decoherence factor magintude} and \eqref{B^2_a_u} lead to the following single-spin expressions:
\begin{align}\nonumber
    |\Gamma^1_{X_0,X'_0}|^2&=1-\delta \xi^2[\sin^2\phi+\sin^2(\tau+\phi)\\\nonumber
&-2\sin\phi\sin(\tau+\phi)\cos(2\tilde{\Delta}\tau)]+O(\xi^4),\\\nonumber
&=\exp\left[-\delta \xi^2\left|\sin\phi+e^{i2\tilde{\Delta}\tau}\sin(\tau+\phi)\right|^2\right]\\\nonumber
&+O(\xi^4)\\\nonumber
B^1_{X_0,X'_0}
    &=1-E(\beta)^2\delta \xi^2[\sin^2\phi+\sin^2(\tau+\phi)\\
&-2\sin\phi\sin(\tau+\phi)\cos(2\tilde{\Delta}\tau)]\\\nonumber
&+O(\xi^4)\\\nonumber
&=\exp\left[-\delta \xi^2E(\beta)^2\left|\sin\phi+e^{i2\tilde{\Delta}\tau}\sin(\tau+\phi)\right|^2\right]\\\nonumber
&+O(\xi^4).
\end{align}
For multiple spins, the law of large numbers can be also used and implies that the exponents above are rescaled by the fraction sizes, similarly to \eqref{Gapprox} and \eqref{Bapprox}:
\begin{align}
& |\Gamma_{X_0,X'_0}|^2 \approx \nonumber\\
&\exp\left[-N_u\frac{\langle g^2\rangle \delta X_0^2}{\Omega^2} \left|\sin\phi+e^{i2\tilde{\Delta}\tau}\sin(\tau+\phi)\right|^2\right], \label{Gapprox2}\\
& B_{X_0,X'_0}\approx \nonumber\\
&\exp\left[-N_{\text{mac}}E(\beta)^2\frac{\langle g^2\rangle \delta X_0^2}{\Omega^2} \left|\sin\phi+e^{i2\tilde{\Delta}\tau}\sin(\tau+\phi)\right|^2\right].\label{Bapprox2}
\end{align}
Figs. \ref{fig:Gamma1_0.5pi} and \ref{fig:B1_0.5pi} clearly show that the periodicity of the objectivity makers for a single macrofraction of spin environment with a non-zero phase, e.g., $\phi=\pm\frac{\pi}{2}$ is broken, though the generalized overlap does not decay much. This will make them decays for multi-spin environment.

We see that the decays are governed by the same length scales \eqref{ldec} and \eqref{ldist} and the functions are doubly periodic with the periods given by $\Omega^{-1}$ and $\Delta^{-1}$.  Apart from this, the behavior in the small $\xi$ approximation 
is essentially the same as for $\phi=0$ case in \eqref{Gapprox} and \eqref{Bapprox}. We will see that this will change for a general $\phi$. Some remarks are in order. For the purpose of this work, we are assuming $\Delta_k$ being the same for all spins. This avoids  complicated averages of the type $\int d\Delta\tanh(\beta\Delta/2)\cos(2\Delta t)$ although obviously different $\Delta_k$ can introduce dephasing,
helping to counter the periodicity. This possibility will be investigated elsewhere. Here, we concentrate on randomized coupling constants $g_k$.

We now estimate the objectivity markers for arbitrary  parameters. According to our procedure [cf. \eqref{LLN1}], we need the averages of the single-spin functions. 
The detailed calculations are presented in Appendix \ref{AppB}.  As before, we separate between the fast and slow oscillating parts:
\begin{align}
\langle|\Gamma^1_{X_0,X'_0}|^2\rangle=\Gamma_{\text{fast}}(\tau)+\Gamma_{\text{slow}}(\tau)\label{Gamma_av_fast_slow}
\end{align}
where oscillating parts are
\begin{align}\nonumber
   \Gamma_{\text{fast}}(\tau)&\equiv\frac{1}{8}\left[2+\sinc\{\delta\bar{\xi}[\sin\phi+\sin(\tau+\phi)]\}\right.\\\nonumber
   &\left.+\sinc\{\delta\bar{\xi}[\sin\phi-\sin(\tau+\phi)]\}\right]\\\nonumber
   &+\frac{E(\beta)^2}{8}\left[2+\sinc\{(\bar{\xi}+\bar{\xi}')\sin\phi+\delta\bar{\xi}\sin(\tau+\phi)]\}\right.\\
   &\left.+\sinc\{(\bar{\xi}+\bar{\xi}')\sin\phi-\delta\bar{\xi}\sin(\tau+\phi)]\}\right]\label{Gamma_fast_phi}
\end{align}
and decaying parts behaving asymptotically as $  1/\sqrt{\tau}$ are 
\begin{align}
    \Gamma_{\text{slow}}(\tau)\equiv\sum_{a,b,c} d^\Gamma_{abc}F^\Gamma[a,b,c]=O\left(\frac{1}{\sqrt{\tau}}\right)\label{Gamma_slow_phi}
\end{align}
with $F^\Gamma[a,b,c]$ defined in (\ref{A:F}) and $d^\Gamma_{abc}$ can be identified in (\ref{A:u0_average}) and (\ref{A:u1_average}). Similarly, 
\begin{align}
    \langle B_{X_0,X'_0}\rangle=B_{\text{fast}}(\tau)+B_{\text{slow}}(\tau)\label{B_av_fast_slow}
\end{align}
where
\begin{align}\nonumber
   B_{\text{fast}}(\tau)&\equiv1-\frac{E(\beta)^2}{8}\left[4-\sinc\{\delta\bar{\xi}[\sin\phi+\sin(\tau+\phi)]\}\right.\\
    &\left.-\sinc\{\delta\bar{\xi}[\sin\phi-\sin(\tau+\phi)\right]\}\\\nonumber
    &\left.-\sinc\{(\bar{\xi}+\bar{\xi}')\sin\phi+\delta\bar{\xi}\sin(\tau+\phi)]\}\right.\\\nonumber
    &\left.-\sinc\{(\bar{\xi}+\bar{\xi}')\sin\phi-\delta\bar{\xi}\sin(\tau+\phi)]\}\right]\label{B_fast_phi}
\end{align}
and
\label{B_slow_phi}\begin{align}
   B_{\text{slow}}(\tau)\equiv\sum_{a,b,c} d^B_{abc}F^B[a,b,c]=O\left(\frac{1}{\sqrt{\tau}}\right)
\end{align}
with $F^B[a,b,c]$ defined in (\ref{A:F}) and $d^B_{abc}$ can be identified in (\ref{A:u0_average}) and (\ref{A:u1_average}). As $\tau\to\infty$, $\Gamma_{\text{slow}}$ and $B_{\text{slow}}$ die out as $1/\sqrt{\tau}$ and only $\Gamma_{\text{fast}}$ and $B_{\text{fast}}$ remain. Thus, as $\tau\to\infty$ a total decoherence factor and a total generalized overlap are given by the fast movers only. 
\begin{align}\nonumber
|\Gamma_{X_0,X'_0}|^2&=[\Gamma_{\text{fast}}(\tau)+\Gamma_{\text{slow}}(\tau)]^{N_u}\\
&=[\Gamma_{\text{fast}}(\tau)]^{N_u}+O\left(\frac{1}{\sqrt{\tau}}\right),\label{Gamma_tot_asym}
\end{align}
\begin{align} \nonumber
 B_{X_0,X'_0}
&=[B_{\text{fast}}(\tau)+B_{\text{slow}}(\tau)]^{N_{\text{mac}}}\\
&=[B_{\text{fast}}(\tau)]^{N_{mac}}+O\left(\frac{1}{\sqrt{\tau}}\right).\label{B_tot_asym}
\end{align}
The trajectory with $\phi=0$, i.e. $X=X_0\cos\tau$ studied before, is rather particular in the structure of $U_{X_0,X'_0}$ [Eq.\eqref{ui_High_Frequency}] in that the slow moving part with the frequency $\tilde{\Delta}(\xi^2-\xi'^2)$ completely vanishes, leaving only the fast moving part in $B_{X_0,X'_0}$. In this oscillatory case,  it is not possible to have a decay of $B_{X_0,X'_0}$ regardless of the number of spins. However, as we see for $\phi\neq0$ the situation is  different as even a small $\phi$ leads to the dephasing of the sinc functions in \eqref{Gamma_fast_phi} and \eqref{B_fast_phi}, which in turn lead to a decay of both functions for many spin environments as we demonstrate below. 

\begin{figure}[t!]
\includegraphics[width=0.9\linewidth, height=4cm]{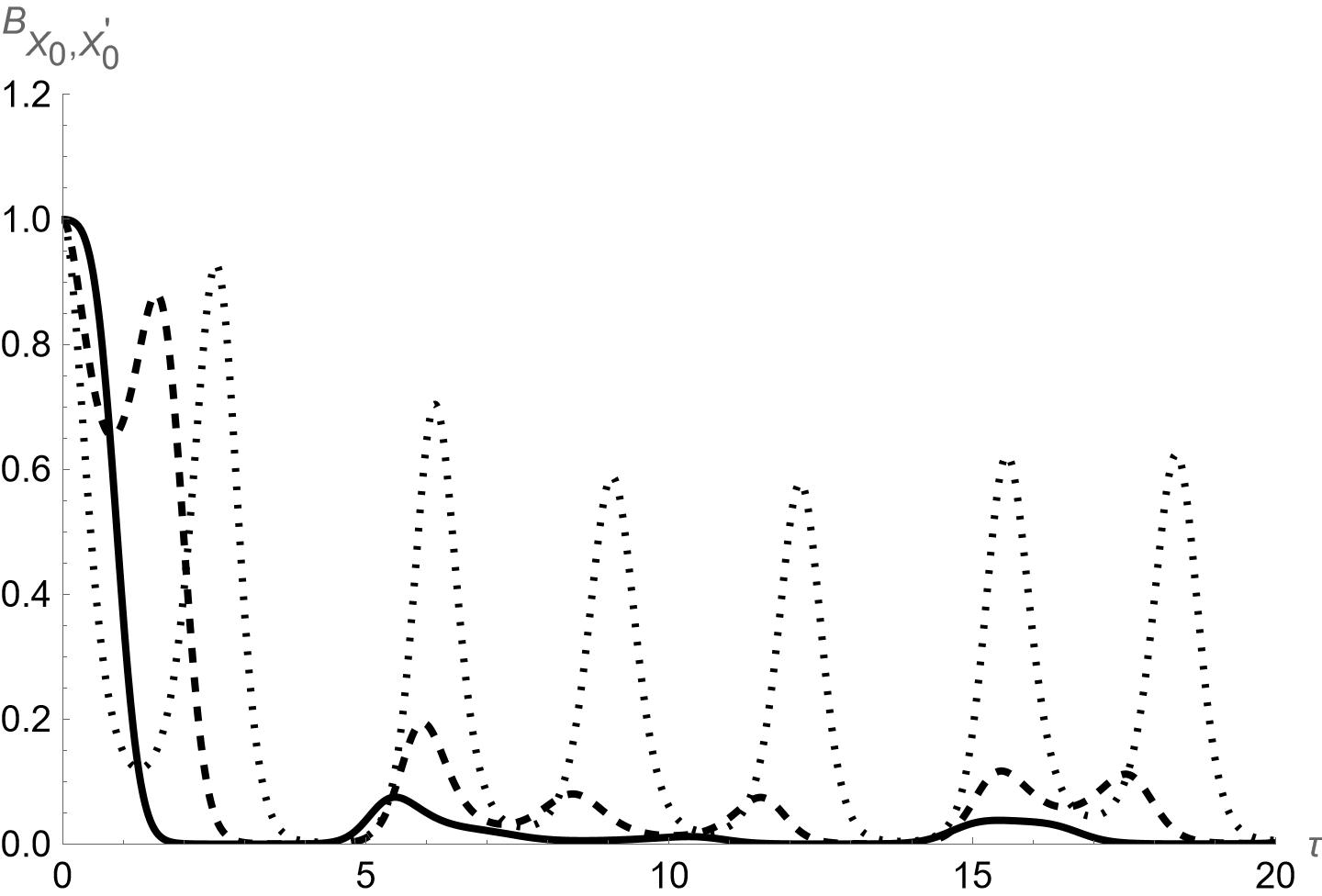} 
\caption{Generalized overlap for large spin environments as a function of time and different phases $\phi$. The dotted, the dashed line, and the thick line correspond to $\phi=\pi/10, \pi/4,\pi/2$, respectively.   $(\bar{\xi}=0.9,\beta\Delta=1,\bar{\xi}'=0.1)$, $N_{mac}=100$ are chosen.)
\label{fig:B_phi}}
\end{figure}

We show that $\max [\Gamma_{\text{fast}}(\tau)]<1$,  $\max[B_{\text{fast}}(\tau)]<1$  for any $\tau>0$, which means both functions are exponentially damped as multiple spins are considered [cf. \eqref{Gamma_tot_asym} and \eqref{B_tot_asym}]. It will be convenient to introduce the following  functions:
\begin{align}
    g_-(\tau)&\equiv\sinc\{\delta\bar{\xi}[\sin\phi+\sin(\tau+\phi)]\}\nonumber\\
    &+\sinc\{\delta\bar{\xi}[\sin\phi-\sin(\tau+\phi)]\}\\
     g_+(\tau)&\equiv\sinc\{(\bar{\xi}+\bar{\xi}')\sin\phi+\delta\xi\sin(\tau+\phi)]\}\nonumber\\
    &+\sinc\{(\bar{\xi}+\bar{\xi}')\sin\phi-\delta\xi\sin(\tau+\phi)]\}.
\end{align}
Their extrema for $\xi,\xi'<1$, which guarantees the High Frequency Expansion and the positivity of $\sinc$ functions, 
are given by:
\begin{align}
  & \sin(\tau+\phi+\pi/2)=0,\\    
   & \sin(\tau+\phi)=0.
\end{align}
Since at $\phi=0$ the second condition  indicates the maxima, $g_\pm(\tau)$ continues being shifted to the left by $\phi\neq0$ as $\sin(\tau+\phi)=0$ moves to the left. So it can be recognized that the first condition gives the minima while the other one the maxima, which read for $t>0$:
\begin{align}\label{Gamma_fast_extrema}
   & \max (\Gamma_{\text{fast}})=\frac{1}{4}\left\{1+\sinc(\delta\bar{\xi}\sin\phi)\right.\\\nonumber
&\left.~~~~~~~~~+E(\beta)^2\left(1+\sinc[(\bar{\xi}+\bar{\xi}')\sin\phi)]\right)\right\}\\
&= \frac{1+ E(\beta)^2}{2}-\frac{\phi^2}{24}\left[\delta\xi^2+ E(\beta)^2 (\bar{\xi}+\bar{\xi}')^2\right]+O(\phi^4)
\end{align}
and
\begin{align}\label{B_fast_extrema}
   & \max (B_{\text{fast}})=1-\frac{E(\beta)^2}{4}\left\{2-\sinc(\delta\bar{\xi}\sin\phi)\right.\nonumber\\
    &\left.~~~~~~~~~~~~~~~~~-\sinc[(\bar{\xi}+\bar{\xi}')\sin\phi)]\right\}\\
    &=1-\frac{\phi^2 }{12}E(\beta)^2(\bar{\xi}^2+\bar{\xi}'^{2})+O(\phi^4)
\end{align}
Note that these  values depend on  $(\bar{\xi},\bar{\xi}')$ and $\beta\Delta$ but not directly on the tunneling energy $\tilde{\Delta}$, which is nevertheless necessary to damp the slow moving parts as we have shown earlier. We see from the above expressions that:
\begin{enumerate}
\item $\max (B_{\text{fast}}) <1$ for any non-zero $\phi$, provided $E(\beta)> 0$, i.e. the temperature is finite, $\beta>0$. 
\item $\max (\Gamma_{\text{fast}})<1$ for any non-zero $\phi$, provided $E(\beta)<1$, i.e. the temperature is non-zero, $\beta<\infty$. 
\end{enumerate}
This, in turn, implies via \eqref{Gamma_tot_asym} and \eqref{B_tot_asym} that outside the temperature extremes, both markers $|\Gamma_{X_0,X'_0}|^2$, $B_{X_0,X'_0}$ are asymptotically damped
for $N_u,N_{\text{mac}} \gg 1 $ and the state \eqref{rho_SoE} approaches the objective state. The amount of damping, and hence the quality of the objectivity in the state,
depends on $N_u$, $N_{\text{mac}}$ and the temperature. For small $\phi$, we obtain from \eqref{Gamma_tot_asym} and \eqref{B_tot_asym} the following bounds:
\begin{align}
& |\Gamma_{X_0,X'_0}|^2 \lessapprox \left[\frac{1+ E(\beta)^2}{2}-\frac{\phi^2}{24}\left[\delta\xi^2+ E(\beta)^2 (\bar{\xi}+\bar{\xi}')^2\right]\right]^{N_u}\nonumber\\
& \approx \left[\frac{1+ E(\beta)^2}{2}\right]^{N_u} \exp\left[-N_u \phi^2\frac{\delta\xi^2+ E(\beta)^2 (\bar{\xi}+\bar{\xi}')^2}{12(1+ E(\beta)^2)}\right]\\
& B_{X_0,X'_0} \lessapprox \left[1-\frac{E(\beta)^2\phi^2 }{12}(\bar{\xi}^2+\bar{\xi}'^{2}) \right]^{N_{\text{mac}}}\nonumber\\
& \approx \exp\left[-N_{\text{mac}}\frac{E(\beta)^2\phi^2 }{12}(\bar{\xi}^2+\bar{\xi}'^{2})\right]
\end{align}
This situation is to be contrasted with the previous Section, where we showed that the cosine trajectory \eqref{traj} does not lead to the permanent damping 
of the generalized overlap for any amount of spins in the macrofraction and hence no permanent objective states can be formed.
Sample plots of both markers, using the exact expressions calculated in Appendix \ref{AppB}, are presented in Figs. \ref{fig:Gamma_phi} and  \ref{fig:B_phi}. 
We see in particular that it is more difficult to damp the generalized overlap as it takes about five times more spins than needed to induce the decoherence. This is to be
expected as the spins are encoding the continuous variable: the oscillation amplitude $X_0$.

\begin{figure}[t!]
\includegraphics[width=0.9\linewidth, height=4cm]{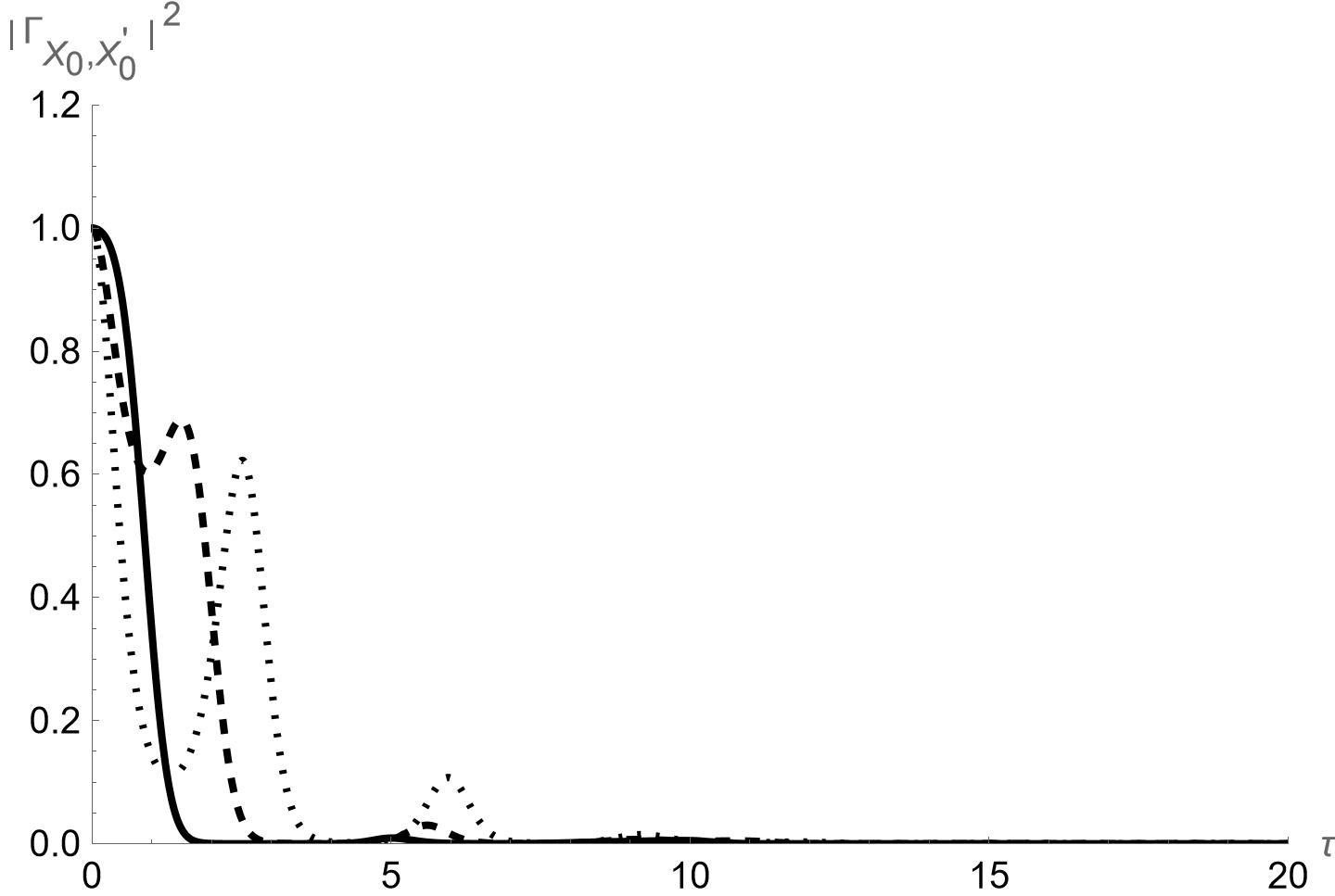} 
\caption{Decoherence factor for large spin environments as a function of time and different phases $\phi$. The dotted, the dashed line, and the thick line correspond to $\phi=\pi/10, \pi/4,\pi/2$, respectively.   $(\bar{\xi}=0.9,\beta\Delta=1,\bar{\xi}'=0.1)$, $N_u=20$ are chosen.)}
\label{fig:Gamma_phi}
\end{figure}

\section{Concluding remarks}
We analyzed the emergence of objectivity in the last canonical models of decoherence, a boson-spin model in SBS state formation that so far has not been used in the rest. Due to the complicated dynamics, we used the recoil-less limit where the influence of the environment on the central oscillator is assumed to be negligible. This is an opposite limit to the usual Born-Markov one, but the most appropriate for studying information transfer to the environment during the decoherence process. The recoil-less limit can be viewed as a version of the Born-Oppenheimer approximation, where the central system evolves unperturbed and affects the environmental spins via coupling to its classical trajectory, which acts as an external time-dependent force. The resulting effective dynamics of the environment allows for the use of the Floquet theory. We perform the analysis in the non-vanishing lowest order of the high-frequency expansion and demonstrate and find in particular, an interesting fact: the fast moving parts of the motion are detrimental to the emergence of objectivity while the slow moving parts enable it. 
Another interesting aspect of the model, not present in other canonical models, is a mismatch between the encoded variable, which is a continuous position-like variable (the oscillation amplitude), and the encoding system, which is finite dimensional (a collection of spins). In this respect, we show two facts. Firstly, we derive two characteristic wavelengths: one corresponding to the decoherence scale on the side of the central system and another governing the resolution, with which collections of environmental spins encode the continuous variable. The lengths are different, in particular, the encoding one depends on the temperature and is larger than the decoherence one, which shows the phenomenon of ``bound information'' in the environment: the resolution of a possible read-out from the environment is lower than the scale on which coherences are destroyed. Both length scales depend on the fraction size, i.e., the bigger the size, the lower the length scales, which is quite intuitive. However, the presence of the length scales does not guarantee a stable formation of objectivity. We show that the latter depends on the type of motion of the central system. In particular, for initial states with a well-defined momentum, there can be only a momentary formation of objectivity, while even a small departure from this specific initial condition, leads to an asymptotic formation of objective states. This is exactly opposite to what one finds in QBM, where initially well-defined momentum states lead to a stable appearance of objectivity, and one of the examples showing how spin and oscillator environments differ.

Our analysis can be applied for the quantum measurement theory in the following points. Firstly, our result is an example of how a continuous variable can be measured by finite-dimensional systems in a realistic scenario of an open quantum dynamics. 
Secondly, from a broader perspective, it could be used to have an exemplary answer to the fundamental question of when the measurement is completed. One may postulate that it is completed when the system $+$ measuring apparatus are close enough to the SBS state, which guarantees an objective character of the measurement result. This remark of course applies to the whole of the SBS and quantum Darwinism program.   

\begin{acknowledgments}
T.H.L. and J.K.K. acknowledge the support
from the Polish National Science Center (NCN), Grant No.  2019/35/B/ST2/01896.
\end{acknowledgments}
\label{appendix}
\appendix

\section{Generalized Overlap in Bloch Representation}\label{AppA}
We will get a geometrical expressions for the generalized overlap $B^2(\rho,\rho')$ for a qubit when $\rho$ and $\rho'$ are unitarily related and its relation with the decoherence factor $|\Gamma_{X_0,X'_0}|^2$. 
$B(\rho,\rho')$ is defined in \eqref{B} as
\begin{align}
   B(\rho,\rho')
   &=\text{Tr}M+2\sqrt{\det M},\label{A:B^2}
\end{align}
where
\begin{align}
M=\sqrt{\rho}\rho'\sqrt{\rho}\label{A:M}
\end{align}
with 
\begin{align*}
  \rho&\equiv  U\rho_0U^\dagger,~\rho'\equiv  U'\rho_0U'^\dagger.
\end{align*}
Using the cyclic property of a trace and a determinant, $B(\rho,\rho')$ in \eqref{A:B^2} is rewritten as
\begin{align}
   B(\rho,\rho')
   &=\text{Tr}\tilde{M}+2\sqrt{\det \tilde{M}},\label{A:B^2_tilde_M}
\end{align}
where $W=U^\dagger U'=U^\dagger_{X_0,X'_0}$ and
\begin{align}
    \tilde{M}=W\rho_0 W^\dagger\rho_0.\label{A:tilde_M}
\end{align}
With the notations $W=u_0-i\vec{u}\cdot\vec{\sigma}$ and $\rho_0=(1+\vec{a}\cdot\vec{\sigma})/2$, we express $W\rho_0$ as
\begin{align}
W\rho_0&=\frac{1}{2}(v_0+\vec{v}\cdot\vec{\sigma}),\label{A:W_rho0}
\end{align}
where
\begin{align}\nonumber
    v_0&\equiv u_0-i\vec{a}\cdot\vec{u}\\
    v_i&\equiv u_0a_i-iu_i+(\vec{u}\times\vec{a})_i.\label{A:w_a,u}
\end{align}
With $\rho_0 W=(W^\dagger\rho_0)^\dagger$ and $W^\dagger(\vec{u}) =W(-\vec{u})$ we obtain $\rho_0 W$:
\begin{align}
\rho_0 W
&=\frac{1}{2}(v_0+\vec{\bar{v}}\cdot\vec{\sigma}),\label{A:rho0_W}
\end{align}
where 
\begin{align}
  \bar{v}_i\equiv u_0a_i-iu_i-(\vec{u}\times\vec{a})_i.\label{A:bar_w_a,u}  
\end{align}
Using \eqref{A:tilde_M},\eqref{A:W_rho0}, and \eqref{A:rho0_W}, we obtain
\begin{align}
   \tilde{M}&= \tilde{m}_0+\sum_i\tilde{m}_i\sigma_i,\label{A:L}
\end{align}
where
\begin{align*}
     \tilde{m}_0&\equiv\frac{1}{4}(|v_0|^2+\vec{v}\cdot\vec{\bar{v}}^*),\\
\tilde{m}_i&\equiv \frac{1}{4}[v_iv^*_0+\bar{v}^*_iv_0+i(\vec{v}\times\vec{\bar{v}}^*)_i].
\end{align*}
and hence $\text{Tr}M$ is expressed as
\begin{align}
\text{Tr}M=\text{Tr}\tilde{M}&=\frac{1}{2}(|v_0|^2+\vec{v}\cdot\vec{\bar{v}}^*).\label{A:TrM}
\end{align}
From \eqref{A:w_a,u} and \eqref{A:bar_w_a,u} 
\begin{align*}
    \sum_iv_i\bar{v}^*_i
    &=u_0^2|\vec{a}|^2-|\vec{u}\times\vec{a}|^2+|\vec{u}|^2,\\
    |v_0|^2&=u^2_0+(\vec{a}\cdot\vec{u})^2,
\end{align*}
Using these $\text{Tr}M$ in \eqref{A:TrM} is expressed in terms of $\vec{a}$ and $\vec{u}$
\begin{align}\nonumber
    \text{Tr}M=2\tilde{m}_0
     &=\frac{1}{2}[1+(2u_0^2-1)|\vec{a}|^2+2(\vec{a}\cdot\vec{u})^2]\\\nonumber
     &=\frac{1}{2}[1+|\vec{a}|^2]-[|\vec{u}|^2|\vec{a}|^2-(\vec{a}\cdot\vec{u})^2]\\
     &=\frac{1}{2}(1+|\vec{a}|^2)-|\vec{a}\times\vec{u}|^2.\label{A:TrM:a,u}
\end{align}
From \eqref{A:M} and $\rho_0=(1+\vec{a}\cdot\vec{\sigma})/2$, $\det M=(\det\rho_0)^2$ is 
\begin{align}
     \det M=\frac{1}{16}(1-|\vec{a}|^2)^2\geq0. \label{A:detL:a}
\end{align}
Plugging $\text{Tr}M$ and $\det M$ in \eqref{A:TrM:a,u} and \eqref{A:detL:a} into \eqref{A:B^2},
\begin{align}
B(\rho,\rho')
   &=1-|\vec{a}\times\vec{u}
   |^2\leq 1.\label{A:B^2:a,u}
\end{align}
With the expression for a decoherence factor $|\Gamma_{X_0,X'_0}|^2=1-|\vec{u}|^2+(\vec{a}\cdot\vec{u})^2$ in \eqref{decoherence factor magintude}, $B(\rho,\rho')$ expression can be related to $|\Gamma_{X_0,X'_0}|^2$,
\begin{align}\nonumber
    B(\rho,\rho')&=(1-|\vec{a}|^2)|\vec{u}|^2+|\Gamma_{X_0,X'_0}|^2\\
    &=(1-|\vec{a}|^2)(1-u^2_0)+|\Gamma_{X_0,X'_0}|^2.\label{A:B^2_Gamma^2}
\end{align}

\section{Objectivity Markers for a  General Classical Trajectory}\label{AppB}
 From (\ref{ui_phi}) it can be seen that the structure of $U_{X_0,X'_0}(\phi)=u_0(\phi)+i\vec{u}(\phi)\cdot\vec{\sigma}$ with a general classical trajectory with arbitrary phase $\phi$ becomes more complicated due to the non-unity initial kick in (\ref{Floquet_U}). For a thermal state (\ref{rthm}), in order to average $|\Gamma_{X_0,X'_0}|^2$ and $B(\rho,\rho')$ over $g$, we need to average $u^2_0(\phi)$ and $u^2_1(\phi)$ over $g$ with a uniform distribution of $g$. The arguments in cosines and sines in (\ref{ui_phi}) are divided into one in the first order in $g$ and one quadratic in $g$. According to (\ref{ui_phi}), we denote the first order arguments by
\begin{align*}
  gk^-&\equiv g(X-X')\sin\phi,\\
   gk^+&\equiv g(X+X')\sin\phi,\\
     gk(\tau)&\equiv g(X-X')\sin(\tau+\phi)
\end{align*}
 and the second order arguments by
\begin{align*}
 s\tau g^2 &\equiv \tilde{\Delta}(X^2-X'^2)\tau g^2,\\
    m\tau g^2+d_\tau&\equiv \tilde{\Delta}(X^2+X'^2)\tau g^2-2\tilde{\Delta}\tau.
\end{align*}
Hence, $U_{X_0,X'_0}$ in (\ref{ui_phi}) is expressed by short notations, $c_{(x)}\equiv\cos x$ and $s_{(x)}\equiv\sin x$:
\begin{align*}
    u_0
    &=c_{(k^-g)}c_{(s\tau g^2)}c_{[k(\tau)g]}+s_{(k^-g)}c_{(m\tau g^2+d_\tau)}s_{[k(\tau) g]},\\
    u_1
    &=-c_{(k^+g)}s_{(sg^2\tau)}c_{[k(\tau)g]}-s_{(k^+g)}s_{(-m\tau g^2-d_\tau)}s_{[k(\tau) g]},\\
     u_2
    &=-s_{(k^+g)}s_{(s\tau g^2)}c_{[k(\tau)g]}+c_{(lg)}s_{(-m\tau g^2-d_\tau)}s_{[k(\tau)g]},\\
     u_3
    &=s_{(k^- g)}c_{(s\tau g^2)}c_{[k(\tau)g]}-c_{(k^-g)}c_{(m\tau g^2+d_\tau)}s_{[k(\tau) g]}.
\end{align*}
$|\Gamma_{X_0.X_0'}|^2$ with the thermal state $\vec{a}=(\tanh[\beta\Delta/2],0,0)$ is expressed as
\begin{align*}
  |\Gamma_{X_0.X_0'}|^2=u^2_0+a^2u^2_1,  
\end{align*}
where
\begin{align}\nonumber
u_0^2&=c^2_{(k^-g)}c^2_{(s\tau g^2)}c^2_{[k(\tau)g]}+s^2_{(k^-g)}c^2_{(m\tau g^2+d_\tau)}s^2_{[k(\tau) g]}\\\nonumber
    &+\frac{1}{2}s_{[2k(\tau)g]}s_{(2k^-g)}c^2_{(sg^2\tau)}c^2_{(m\tau g^2+d)},\\\nonumber
u^2_1&=c^2_{(k^+g)}s^2_{(sg^2\tau)}c^2_{[k(\tau)g]}+s^2_{(k^+g)}s^2_{(-m\tau g^2-d_\tau)}s^2_{[k(\tau) g]}\\
    &-\frac{1}{2}s_{[2k(\tau)g]}s_{(2k^+g)}s_{(sg^2\tau)}s_{(-m\tau g^2-d_\tau)}\label{A:u1,u2}
\end{align}
Products of cosines and sines are decomposed into single cosines and sines. For instance, the first term in $u_0^2$ in (\ref{A:u1,u2}) is decomposed into cosine functions
\begin{align*}
   &c^2_{(k^-g)}c^2_{(s\tau g^2)}c^2_{[k(\tau)g]}\\
   &= \frac{1}{8}+\frac{1}{8}\{c_{(2k^-g)}+c_{(2s\tau g^2)}+c_{[2k(\tau)g]}\}\\
    &+\frac{1}{16}\{c_{(2s\tau g^2+2k^-g)}+c_{(2s\tau g^2-2k^-g)}+c_{[2s\tau g^2+2k(\tau) g]}\\
    &+c_{[2s\tau g^2-2k(\tau) g]}+c_{[2k(\tau) g+2k^-g]}+c_{[2k(\tau) g-2k^-g]}\}\\
    &+\frac{1}{32}\{c_{(s\tau g^2+2k(\tau) g+k^-g)}+c_{[s\tau g^2+2k(\tau) g-k^-g]}\\
    &+c_{(s\tau g^2-2k(\tau) g+k^-g)}+c_{[s\tau g^2-2k(\tau) g-k^-g]}\}.
\end{align*}
All terms in \eqref{A:u1,u2} are finally into cosine functions of linear or quadratic functions of $g$.
We define an average value as an integral of a function over $g$ from 0 to a maximum value $\bar{g}$,
\begin{align*}
    \langle f(g) \rangle\equiv\frac{1}{\bar{g}}\int^{\bar{g}}_0 dg f(g).
\end{align*}
Relevant functions to consider are only a cosine function with different arguments. For quadratic arguments of cosine $\langle\cos[ag^2+bg+c]\rangle$ for $a\neq0$ is
\begin{align*}
    &\frac{1}{\bar{g}}\int^{\bar{g}}_0dg\cos[ag^2+bg+c]\\
    &=\frac{1}{\sqrt{a}\bar{g}}\cos\left(\frac{b^2}{4a}-c\right)\left[C\left(\sqrt{a}\bar{g}+\frac{b}{2\sqrt{a}}\right)-C\left(\frac{b}{2\sqrt{a}}\right)\right]\\
    &+\frac{1}{\sqrt{a}\bar{g}}\sin\left(\frac{b^2}{4a}-c\right)\left[S\left(\sqrt{a}\bar{g}+\frac{b}{2\sqrt{a}}\right)-S\left(\frac{b}{2\sqrt{a}}\right)\right],
\end{align*}
where $C(x)\equiv \int_0^x du \sin u^2$ and $S(x)\equiv \int_0^x du \sin u^2$. 
For linear arguments,  $\langle \cos(bg +c)\rangle$ for $b\neq0$ is
\begin{align*}
    \frac{1}{\bar{g}}\int^{\bar{g}}_0dg\cos[bg+c]&=\frac{1}{\bar{g}b}(\sin[b\bar{g}+c]-\sin[c]),
\end{align*}
Especially, for instance, the average value of $2\cos Ag^2\cos Bg$ is given by
\begin{align}\nonumber
&2\langle \cos Ag^2\cos Bg\rangle = \langle \cos(Ag^2+Bg)\rangle + \langle \cos(Ag^2-Bg)\rangle\nonumber\\\nonumber
&=\cos\left(B^2/4A\right) \left\langle\cos[A(g+B/2A)^2] + \cos[A(g-B/2A)^2]\right\rangle\\\nonumber
&+\sin\left(B^2/4A\right) \left\langle\sin[A(g+B/2A)^2] + \sin[A(g-B/2A)^2]\right\rangle\\\nonumber
&=\frac{\cos\left(B^2/4A\right) }{\sqrt{ \bar g^2 A}}\left\{C[\sqrt{ A}(\bar g +B/2A)] + C[\sqrt{ A}(\bar g -B/2A)] \right\}\\
& +\frac{\sin\left(B^2/4A\right) }{\sqrt{ \bar g^2 A}}\left\{S[\sqrt{ A}(\bar g +B/2A)] + S[\sqrt{ A}(\bar g -B/2A)] \right\}\label{A:slow4}
\end{align}
Finally, average values of $u^2_0$ and $u^2_1$ in (\ref{A:u1,u2}) are written in terms of constants, sinc functions and the Fresnel pairs $F[a,b,c]$:
\begin{align}\label{A:u0_average}\nonumber
    \langle u^2_0\rangle&=\frac{1}{4}+\frac{1}{8}\sinc[\omega^-+\omega(\tau)]+\frac{1}{8}\sinc[\omega^--\omega(\tau)]\\\nonumber
   &+\frac{1}{16}F[s\tau,0,0]+\frac{1}{16}F[m\tau,0,d_\tau]\\\nonumber
   &+\frac{1}{16}F[s\tau,\omega^-,0]+\frac{1}{16}F[s\tau,\omega(\tau),0]\\\nonumber
    &+\frac{1}{32}F[s\tau,\omega^-+\omega(\tau),0]+\frac{1}{32}F[s\tau,\omega^--\omega(\tau),0]\\\nonumber
    &-\frac{1}{16}F[m\tau,\omega^-,d_\tau]-\frac{1}{16}F[m\tau,\omega(\tau),d_\tau]\\\nonumber
    &+\frac{1}{32}F[m\tau,\omega^-+\omega(\tau),d_\tau]+\frac{1}{32}F[m\tau,\omega^--\omega(\tau),d_\tau]\\\nonumber
    &+\frac{1}{16}F[(m+s)\tau/2,\omega^--\omega(\tau),d_\tau/2]\\
    &-\frac{1}{16}F[(m+s)\tau/2, \omega^-+\omega(\tau), d_\tau/2]\\\nonumber
    &+\frac{1}{16}F[(m-s)\tau/2,\omega^--\omega(\tau),d_\tau/2]\\\nonumber
    &-\frac{1}{16}F[(m-s)\tau/2, \omega^-+\omega(\tau), d_\tau/2]
\end{align}
and
\begin{align}\label{A:u1_average}\nonumber
    \langle u^2_1\rangle
    &=\frac{1}{4}+\frac{1}{8}\sinc[\omega^++\omega(\tau)]+\frac{1}{8}\sinc[\omega^+-\omega(\tau)])\\\nonumber
    &-\frac{1}{16}F[s\tau,0,0]-\frac{1}{16}F[m\tau, 0, d_\tau ]\\\nonumber
    &-\frac{1}{16}F[s\tau,\omega^+,0]-\frac{1}{16}F[s\tau,\omega(\tau),0]\\\nonumber
    &-\frac{1}{32}F[s\tau,\omega^++\omega(\tau),0]-\frac{1}{32}F[s\tau,\omega^+-\omega(\tau),0]\\\nonumber
&+\frac{1}{16}F [m\tau, \omega^+, d_\tau] + \frac{1}{16}F [m\tau, \omega(\tau), d_\tau]\\\nonumber
&-\frac{1}{32}F [m\tau, \omega^++\omega(\tau),d_\tau]-\frac{1}{32}F[m\tau, \omega^+-\omega(\tau), d_\tau] \\\nonumber
    &+\frac{1}{16}F[(m+s)\tau/2,\omega^+-\omega(\tau),d_\tau/2]\\\nonumber
    &-\frac{1}{16}F[(m+s)\tau/2,\omega^++\omega(\tau),d_\tau/2]\\
    &-\frac{1}{16}F[(m-s)\tau/2,\omega^+-\omega(\tau),d_\tau/2]\\\nonumber
    &+\frac{1}{16}F[(m-s)\tau/2,\omega^++\omega(\tau),d_\tau/2],
\end{align}
where $\xi\equiv\bar{g}X$, $\xi'\equiv\bar{g}X'$, 
\begin{align*}
  &\omega(\tau)\equiv2\delta\xi\sin(\tau+\phi),\\
    &~\omega^-=2\delta\xi\sin\phi,~\omega^+=2(\xi+\xi')\sin\phi,   
\end{align*}
\begin{align}\nonumber
    &F[a,b,c]\equiv\\\nonumber
    &=\langle \cos(a g^2+bg +c)\rangle+\langle \cos(a g^2-bg +c)\rangle\\\nonumber
     &=\frac{1}{\sqrt{a}\bar{g}}\cos\left(\frac{b^2}{4a}-c\right)\left[C\left(\sqrt{a}\bar{g}+\frac{b}{2\sqrt{a}}\right)\right.\\\nonumber
     &~~~~~~~~~~~~~~~~~~~~~~~~~~~~~\left.+C\left(\sqrt{a}\bar{g}-\frac{b}{2\sqrt{a}}\right)\right]\\\nonumber
    &+\frac{1}{\sqrt{a}\bar{g}}\sin\left(\frac{b^2}{4a}-c\right)\left[S\left(\sqrt{a}\bar{g}+\frac{b}{2\sqrt{a}}\right)\right.\\
    &~~~~~~~~~~~~~~~~~~~~~~~~~~~~~\left.+S\left(\sqrt{a}\bar{g}-\frac{b}{2\sqrt{a}}\right)\right]\label{A:F}
\end{align}
and
    \begin{align*}
    \sinc[b]&\equiv\frac{1}{\bar{g}}\int^{\bar{g}}_0dg\cos[bg]=\frac{\sin[b\bar{g}]}{b\bar{g}}.
\end{align*}

\end{document}